\documentstyle[11pt,aaspp4,flushrt,psfig]{article}
\journalid{VOL}{JOURNAL DATE 1997}
\articleid{START PAGE}{END PAGE}
\newcommand\bff{\rm}
\newcommand\itt{\rm}
\def\aaDCE#1#2#3{ 19#1, {\itt A.A.,} {\bff #2}, #3}

\def\apjDCE#1#2#3{ 19#1, {\itt Ap.J.,} {\bff #2}, #3}
\def\apjletDCE#1#2#3{ 19#1, {\itt Ap.J.(Letts),} {\bff  #2}, #3}
\def\apjpressDCE{{\itt Ap. J.,} in press}

\def\apjsDCE#1#2#3{ 19#1, {\itt Ap.J.Suppl.,} {\bff #2}, #3}

\def\appDCE#1#2#3{ 19#1, {\itt Astroparticle Phys.,} {\bff #2}, #3}
\def\astroletsDCE#1#2#3{ 19#1, {\itt Astronomy Letters.,} {\bff #2}, #3}

\def\icrcplovdivDCE#1#2{ 1977, in {\itt Proc. 15th ICRC (Plovdiv)},
   {\bff #1}, #2}
\def\icrcromeDCE#1#2{ 1995, in {\itt Proc. 24th ICRC(Rome)},
  {\bff #1}, #2}
\def\icrcsaltlakeDCE#1#2{ 1999, {\itt Proc. 26th Int. Cosmic Ray Conf.
    (Salt Lake City),} {\bff #1}, #2}
\def\icrcsaltlakeADCE#1#2{ 1999a, {\itt Proc. 26th Int. Cosmic Ray Conf.
    (Salt Lake City),} {\bff #1}, #2}
\def\icrcsaltlakeBDCE#1#2{ 1999b, {\itt Proc. 26th Int. Cosmic Ray Conf.
    (Salt Lake City),} {\bff #1}, #2}

\def\JETPDCE#1#2#3{ 19#1, {\itt JETP, } {\bff #2}, #3}

\def\jgrDCE#1#2#3{ 19#1, {\itt J.G.R., } {\bff #2}, #3}
\def\jgrpress2000DCE{ 2000, {\itt J.G.R.}, in press}

\def\mnrasDCE#1#2#3{ 19#1, {\itt M.N.R.A.S.,} {\bff #2}, #3}
\def\natureDCE#1#2#3{ 19#1, {\itt Nature,} {\bff #2}, #3}

\def\phyreptsDCE#1#2#3{ 19#1, {\itt Phys. Repts.,} {\bff #2}, #3}

\def\rppDCE#1#2#3{ 19#1, {\itt Rep. Prog. Phys.,} {\bff #2}, #3}

\def\spuDCE#1#2#3{ 19#1, {\itt Sov. Phys. Usp.,} {\bff #2}, #3}
\def\ssrDCE#1#2#3{ 19#1, {\itt Space Sci. Rev.,} {\bff #2}, #3}
\newcommand{\pmb}[1]{#1{\setbox0=\hbox{#1}%
  \kern-0.025em\copy0\kern-\wd0
  \kern0.05em\copy0\kern-\wd0
  \kern-0.025em\raise0.0433em\box0 }}
\newcount\llistno
\newcommand\llist{\global\advance \llistno by 1 {(\the\llistno) }}
\newcommand\newlist{\llistno=0}
\newcount\Inum
\newcount\IInum
\newcount\IIInum
\newcount\IVnum
\newcommand\I{\global\multiply\IInum by 0 \global\multiply\IIInum by 0
            \global\multiply\IVnum by 0 \global\advance \Inum by 1
            {\the\Inum. }}
\newcommand\II{\global\multiply\IIInum by 0\global\multiply\IVnum by 0
       \global\advance \IInum by 1 {\the\Inum.\the\IInum. }}
\newcommand\III{\global\multiply\IVnum by 0\global\advance \IIInum by 1
            {\the\Inum.\the\IInum.\the\IIInum. }}
\newcommand\IV{\global\advance \IVnum by 1
            {\the\IVnum. }}
\newcommand\back{\leftskip0pt}
\newcommand\Fnu{F_{\nu}}
\newcommand\Rphot{R_{\rm phot}}
\newcommand\ProDenUpS{n_{p0}}

\newcommand\BEname{Berezhko \& Ellison}
\newcommand\BKP{Berezhko, Ksenofontov, \& Petukhov} 
\newcommand\EnNR{E_{\rm NR}}
\newcommand\Enrel{E_{\rm rel}}
\newcommand\RH{Rankine-Hugoniot}
\def\xx#1{\!\times\!10^{#1}}
\newcommand\EffRel{\epsilon_{\mathrm{rel}}}

\newcommand\gamgas{\gamma_{\mathrm{2g}}}
\newcommand\engam{\displaystyle{\varepsilon}_{\gamma}}
\newcommand\entot{\gamma_{\rm e}}
\newcommand\TempPro{T_{p0}}
\newcommand\TempRatio{T_{\mathrm{e2}}/T_{\mathrm{p2}}}
\newcommand\DStemp{T_{\mathrm{p2}}}
\newcommand\DStp{T_{\mathrm{tp}}}
\newcommand\epRatio{(e/p)_{\mathrm{rel}}}

\newcommand\Bcbr{B_{\mathrm{cbr}}}

\newcommand\EnSN{E_{\mathrm{sn}}}
\newcommand\Mej{M_{\mathrm{ej}}}
\newcommand\tSNR{t_{\mathrm{snr}}}
\newcommand\dSNR{D_{\mathrm{snr}}}
\newcommand\EmisVol{V_{\mathrm{emis}}}
\newcommand\Vsk{V_{\mathrm{sk}}}
\newcommand\Rsk{R_{\mathrm{sk}}}
\newcommand\Rpiston{R_{\mathrm{piston}}}
\newcommand\rhoShell{\rho_{\mathrm{shell}}}
\newcommand\tacc{t_{\mathrm{acc}}}
\newcommand{\MAZ}{M_{\mathrm{A0}}}
\newcommand{\MSZ}{M_{\mathrm{S0}}}
\newcommand\pcc{cm$^{-3}$}
\newcommand\kmps{km s$^{-1}$}
\newcommand\muG{$\mu$G}
\newcommand\etainjP{\eta_{\mathrm{inj,p}}}
\newcommand\etainjE{\eta_{\mathrm{inj,e}}}
\newcommand\etamfp{\eta_{\mathrm{mfp}}}
\newcommand\etamfpElec{\eta_{\mathrm{mfp,e}}}
\newcommand\etamfpPro{\eta_{\mathrm{mfp,p}}}
\newcommand\rg{r_{\mathrm{g}}}
\newcommand\rgmax{r_{\mathrm{g,max}}}

\newcommand\mpc{m_{\mathrm{p}} \, c}
\newcommand\amc{a_{\mathrm{mc}}}
\newcommand\ainjE{a_{\mathrm{inj,e}}}
\newcommand\ainjP{a_{\mathrm{inj,p}}}
\newcommand\ainj{a_{\mathrm{inj}}}
\newcommand\amax{a_{\mathrm{max}}}
\newcommand\amaxe{a_{\mathrm{max,e}}}
\newcommand\amaxp{a_{\mathrm{max,p}}}
\newcommand\qsub{q_{\mathrm{sub}}}
\newcommand\qint{q_{\mathrm{int}}}
\newcommand\qmin{q_{\mathrm{min}}}
\newcommand\qTP{q_{\mathrm{tp}}}
\newcommand\pinj{p_{\mathrm{inj}}}
\newcommand\pinjp{p_{\mathrm{inj,p}}}
\newcommand\pinje{p_{\mathrm{inj,e}}}
\newcommand\pinterm{p_{\mathrm{01}}}
\newcommand\pmax{p_{\mathrm{max}}}
\newcommand\Rsub{r_{\mathrm{sub}}}
\newcommand\Rtot{r_{\mathrm{tot}}}
\newcommand\Emax{E_{\mathrm{max}}}
\newcommand\EmaxPro{E_{\mathrm{max,p}}}
\newcommand\EmaxElec{E_{\mathrm{max,e}}}
\newcommand\MC{Monte Carlo}
\newcommand\DSA{diffusive shock acceleration}
\newcommand\iec{i.e., }
\newcommand\egc{e.g., }
\newcommand\etal{et al.}
\newcommand\brem{bremsstrahlung}
\newcommand\brems{bremsstrahlung}
\newcommand\syn{synchrotron}
\newcommand\synch{synchrotron}
\newcommand\IC{inverse-Compton}
\newcommand\pion{pion-decay}
\newcommand\alf{Alfv\'en}

\begin{document}
\newcommand{\figureout}[2]{\centerline{}
   \centerline{\psfig{figure=#1,width=5.5in}}
    \figcaption{#2}\clearpage }
\newcommand{\figureoutsmall}[2]{\centerline{\psfig{figure=#1,width=5.0in}}
    \figcaption{#2}\clearpage }
\newcommand{\figureoutvsmall}[2]{\centerline{\psfig{figure=#1,width=4.0in}}
    \figcaption{#2}\clearpage }

\vphantom{p}
\vskip -40pt 
{\it Astrophysical Journal}, submitted, January 2000; in press March 2000
\vskip 15pt
\title{NONLINEAR SHOCK ACCELERATION\\
       AND PHOTON EMISSION IN SUPERNOVA REMNANTS}

   \author{Donald C. Ellison}
   \affil{Department of Physics, North Carolina State University,\\
          Box 8202, Raleigh NC 27695, U.S.A.\\
          \it don\_ellison@ncsu.edu\rm}

   \author{Evgeny G. Berezhko}
   \affil{Institute of Cosmophysical Research and Aeronomy,\\
          Lenin Avenue 31, 677891 Yakutsk, Russia\\
          \it berezhko@sci.yakutia.ru\rm}

   \and

   \author{Matthew G. Baring\altaffilmark{1}}
   \affil{Laboratory for High Energy Astrophysics, Code 661, \\
      NASA Goddard Space Flight Center, Greenbelt, MD 20771, U.S.A.\\
      \it baring@lheavx.gsfc.nasa.gov\rm}

   \altaffiltext{1}{Universities Space Research Association}
%


\begin{abstract}
We have extended a simple model of nonlinear
\DSA\ (Berezhko \& Ellison \cite{BEapj99}; 
Ellison \& Berezhko \cite{EBicrc99a}) 
to include the injection and acceleration of electrons
and the production of photons from \brem, \syn, \IC, and \pion\
processes.  We argue that the results of this model, which is simpler
to use than more elaborate ones, offer a significant improvement over
test-particle, power-law spectra which are often used in astrophysical
applications of \DSA. With an evolutionary supernova remnant (SNR)
model to obtain shock parameters as functions of ambient interstellar
medium parameters and time, we predict broad-band continuum photon
emission from supernova remnants in general, and SN1006 in particular,
showing that our results compare well with the more complete
time-dependent and spherically symmetric nonlinear model of 
\BKP\ (\cite{BereKP99a}). 
We discuss the implications nonlinear shock acceleration has for X-ray
line emission, and use our model to describe how ambient conditions
determine the TeV/radio flux ratio, an important parameter for
$\gamma$-ray observations of radio SNRs.
\end{abstract}
\clearpage

\section{INTRODUCTION}
 \label{sec:intro}

More than twenty years of spacecraft observations in the heliosphere
have proven that collisionless shocks can accelerate particles with
high efficiency, \iec 10-50\% of the ram energy can go into superthermal
particles (\egc Eichler \cite{Eich81}; Gosling \etal\
\cite{GoslingEtal81}; 
Ellison \etal\ \cite{EMP90}). 
A recent example comes from Terasawa \etal\ (\cite{Terasawa99})
who
report on an interplanetary shock, observed by GEOTAIL, where the
pressures in thermal and accelerated particles, and the magnetic field
are in approximate equipartition in the foreshock. They observed that
the shock transition was smoothed by the energetic particle
backpressure, giving unambiguous evidence that nonlinear acceleration
occurred in an interplanetary shock, weak by astrophysical standards
(for this exceptionally strong interplanetary shock, the sonic Mach
number was $\MSZ \sim 5-10$).
Energetic particles exist throughout the
universe and shocks are commonly associated with them, confirming that
shock acceleration is important beyond the heliosphere.  In
fact, shocks are believed to be the main source of Galactic cosmic
rays, and energetic particles from shocks may produce spectacular
events such as $\gamma$-ray bursts and 
X-ray and gamma-ray flaring in blazars.  
The shocks in these objects are expected to be much stronger than
those in the heliosphere and can only be more efficient and
nonlinear.
The conjecture that collisionless shocks are efficient accelerators is
strengthened by results from plasma simulations which show efficient
shock acceleration consistent with spacecraft observations (\egc
Scholer, Trattner, \& Kucharek \cite{STK92}; Giacalone \etal\
\cite{GBSE97}).  Other indirect evidence comes from the implied
efficiency of galactic cosmic ray production, radio emission from
supernova remnants (SNRs) (see Reynolds \& Ellison \cite{RE92}),
equipartition arguments in AGNs and $\gamma$-ray bursts, etc. (see
Blandford \& Eichler \cite{BE87} for an early review). There is also clear
evidence that shocks can produce strong self-generated
turbulence. This has long been seen in heliospheric shocks (\egc Lee
\cite{Lee82}, \cite{Lee83}; Kennel \etal\ \cite{KennelEtal84}; 
Baring \etal\ \cite{BaringEtal97}) and there is
evidence that it occurs at SNRs as well (\iec Achterberg, Blandford,
\& Reynolds \cite{ABR94}).

Despite this compelling evidence for efficient, nonlinear shock
acceleration, many astrophysical applications of shock
acceleration continue to
use the test-particle theory proposed more than 20 years ago by
Axford, Leer, \& Skadron (\cite{ALS77}), Krymskii (\cite{Krym77}),
Bell (\cite{Bell78}), and Blandford \& Ostriker (\cite{BO78}).  We
believe it is possible that test-particle models are used in
situations where nonlinear shocks are clearly expected because the
numerical, nonlinear results are complicated and unwieldy.  None of
the current techniques modeling nonlinear shocks have simple analytic
approximations and this makes it difficult to perform parameter
searches or compare nonlinear results to observations.

Here, in an attempt to remedy this situation, we present a simple
model of \DSA, based on more complete studies, which includes the
essential physics of nonlinear acceleration when the backpressure from
energetic ions modifies the shock structure and induces spectral
curvature.  This model is computationally fast and easy-to-use,
but complete enough to be a valuable tool for interpreting
observations.  We also investigate some implications of efficient
acceleration and the associated nonlinear effects on the modeling and
interpretation of particle and photon observations of SNRs, and
include a detailed study of the broad-band continuum emission from the
forward shock in SN1006.

Our basic model is given in \BEname\ (\cite{BEapj99}) where only
protons are considered. Here, we extend the model to include electrons
and alpha particles, and calculate the broad-band, continuum photon
emission from the ion and electron distributions (Ellison \& Berezhko
\cite{EBicrc99a}).  For protons, the injection process whereby some
fraction of the shock heated plasma becomes accelerated to
superthermal energies is, of necessity, parameterized but the model
allows for the investigation of parameters against observations.
Unfortunately, the theory of electron injection and acceleration in
shocks is on a less secure footing than for protons, so we are forced
to use additional parameters for electron injection. Helium is
included without accounting for the enhancement effect known to occur
for high mass to charge number ions (\egc Baring
\etal\ \cite{BaringEtal99}), 
but this approximation has little effect on the results presented
here.

We claim our nonlinear model, in spite of its approximations and
parameters, is far more physically meaningful than the test-particle
power laws that are still routinely assumed by many workers as the
outcome from shock acceleration.  A crucial property
of efficient, nonlinear shock acceleration is the interconnection of
the entire particle distribution from thermal to the highest energies,
and the linkage between protons and electrons. Because energy is
conserved, a change in the production efficiency of the highest energy
cosmic rays {\it must} impact the thermal properties of the shock
heated gas and vice versa. If more energy goes into relativistic
particles, less is available to heat the gas. In contrast, the power
laws assumed by test-particle models have no connection with the
thermal gas, energy conservation does not constrain the normalization
of the power law, and the spectral index can be changed with no
feedback on the thermal plasma.  The availability of an easy-to-use
nonlinear shock acceleration model will have important implications
for interpreting the broad-band emission from cosmic sources.  In
particular, the model presented here can efficiently explore parameter
space to determine which quantities are the most dominant in
controlling the spectral character and detectability of a remnant's
emission in each waveband.  The density of the ambient interstellar
medium and the environmental magnetic field strength are the most
interesting and critical parameters in this regard.  It is clear that
adjusting parameters to fit one component, say radio, changes the
predicted fluxes at all other frequencies and most significantly in
the X-ray band.  Based on constraints imposed by radio and
$\gamma$-ray observations, the simple model can distinguish, as a
function of source model and environmental parameters, the relative
contributions of \synch\ (from TeV electrons) and non-thermal
\brems\ to the X-ray continuum.  Our model predicts the shape of the
full, nonthermal electron spectrum and, for sources where non-thermal
\brems\ is significant, X-ray line models may need to consider these
non-thermal distributions.

Conversely, the model possesses the ability to predict TeV fluxes
given detections at other energies.  One of the most perplexing
developments of cosmic ray physics is the lack of clear detections of
pion-decay photons from SNRs.  The original predictions of Drury,
Aharonian, \& V\"olk (\cite{DAV94}) are well above current upper
limits from EGRET and ground-based TeV telescopes. It is almost
universally assumed that supernova blast waves accelerate cosmic ray
ions at least up to the ``knee'' near $10^{15}$ eV, and if so, these
high energy ions will interact with the ambient gas and produce pions
which will decay into GeV and TeV photons. The distinctive pion-decay
bump was expected to be seen with current sensitivities and the lack
of detection means there is still no direct evidence that SNRs produce
cosmic ray ions and is something of a concern for both theorists and
builders of $\gamma$-ray telescopes.  Since the detectability depends
on source parameters, observers critically need selection criteria
which reliably predict which SNRs, observed at other frequencies, are
likely to be bright at $\gamma$-ray energies.  This is particularly
true for ground-based air \v{C}erenkov telescopes which must devote
weeks or even months observing a single position in the sky to obtain
good statistics.  As we show in our survey below, the model can
provide these selection criteria. For most of the cases we have
studied, supernov\ae\ which explode in low-density, low magnetic field
regions have the largest TeV to radio flux ratios.

\section{SIMPLE NONLINEAR SHOCK MODEL}
 \label{sec:simple}

The model of Berezhko \& Ellison
(\cite{BEapj99})  synthesizes the essential features of two
complimentary approaches to nonlinear shock acceleration, namely the
semi-analytical diffusion equation method of Berezhko et
al. (\cite{BerezEtal96}) and others (\egc Kang \& Jones \cite{KJ91},
\cite{KJ95}) and the Monte Carlo technique of \egc Ellison \etal\
(\cite{EBJ96}).  The premise of the model of Berezhko \& Ellison is
that the main aspects of time-dependent, nonlinear shock acceleration
can be understood in the framework of a plane-wave, steady-state
assumption with simple approximations for the accelerated particle
spectrum.  An essential element of nonlinear acceleration is particle
escape and Ellison \& Berezhko (\cite{EBicrc99b}) showed (and we
confirm here) that the
dynamic effects of particle escape are essentially the same in a
plane-wave, steady-state shock as in an evolving, spherical shock.
The two very different scenarios agree because the process of particle
decoupling (\iec escape) from shocks at high energies produces the
same shock modification and increased compression ratio in the
evolving, spherical shock solution, where particle `escape' is
mimicked by energy `dilution' in the ever expanding upstream volume,
as it does in the plane-wave approximation, where particles leave at a
maximum cutoff energy or a free-escape-boundary (FEB). The essential
point is that particle escape from strong shocks is a {\it fundamental
part} of the nonlinear acceleration process and is determined
primarily by energy and momentum conservation, not time-dependence or
a particular geometry.  The injection efficiency, together with the
`shock size' (however it may be determined), set the shock structure
and determine the overall acceleration efficiency.

In other words, in a real, finite sized shock, geometry is
important for determining the maximum value of momentum, $\pmax$,
where the spectrum cuts off, and particle escape comes about
naturally. In the plane-wave approximation, escape is parameterized
with a FEB, but the model can still be used to good effect with this
restriction.  Similarly, since the majority of particles accelerated
by an expanding SNR blast wave typically have acceleration times
which are much shorter than the time scale of the system (in fact,
even the highest energy particles expected in SNRs have acceleration
times a few times shorter than the remnant age, \egc Berezhko
\cite{Berez96}), a steady-state approximation can be used for modeling
the particles accelerated by a strongly modified shock.  We give a
quantitative verification of this claim in Ellison \& Berezhko
(\cite{EBicrc99b}), and demonstrate it below in detail for the SNR
SN1006.

\subsection{Protons}
 \label{sec:simple_p}

The non-linearity of the acceleration process in high Mach number
shocks is manifested through the feedback of the ions on the
spatial profile of the flow velocity (\egc Drury \cite{Drury83}; Jones
\& Ellison \cite{JE91}),
which in turn determines the particle distribution.  The accelerated
population presses on the upstream plasma and slows it.  An upstream
precursor forms, in which the flow speed (in the absence of
instabilities) is monotonically decreasing.  The net effect that
emerges is one where the overall compression ratio, $\Rtot$, from far
upstream to far downstream of the subshock discontinuity, exceeds that
obtained in the test-particle scenario, while the subshock compression
ratio, $\Rsub$, which is mainly responsible for heating the gas, is
less than the test-particle compression.  This phenomenon was
identified by Eichler (\cite{Eich84}), and Ellison \& Eichler
(\cite{EE84}), and arises for two reasons. The most important is that
particle escape from strong shocks drains energy and pressure which
must be compensated for by ramping up the overall compression ratio to
conserve the fluxes. The second reason is that relativistic particles,
with their softer equation of state, contribute significantly to the
total pressure making the shocked plasma more compressible.  Since
particle diffusion lengths are generally increasing functions of
momentum (\egc Giacalone \etal\
\cite{GBSE93}; Scholer, Kucharek, \& Giacalone \cite{SKG2000}), 
high momentum particles sample a broader portion of the flow velocity
profile, and hence experience larger compression ratios than low
momentum particles. Consequently, higher momentum particles have a
flatter power-law index than those at lower momenta, thereby
dominating the pressure in a nonlinear fashion and producing a
concave upward spectral curvature which is the trademark of nonlinear
shock acceleration.

It is precisely this spectral concavity that can be compactly
approximated by the simple model of Berezhko \& Ellison (\cite{BEapj99}).
\placefigure{fig:EBfive}              
Berezhko \& Ellison assume that the accelerated part
of the shocked particle phase-space distribution, $f(p)$, above a
superthermal injection momentum, $\pinj$, can be described as a
three-component power law:
\begin{equation}
\label{eq:ThreePower}
f(p)=\cases{
 \ainj \left( p / \pinj \right)^{-\qsub}
 &if \ \  $\pinj \le p \le \mpc $
\ ,
\cr
\noalign{\kern5pt}
\amc \left [ p / (\mpc) \right]^{-\qint}
&if \ \ $ \mpc  \le p\le \pinterm $
\ ,
\cr
\noalign{\kern5pt}
\amax \left( p / \pinterm  \right )^{-\qmin}
&if \ \ $ \pinterm \le p \le \pmax$
\ .
\cr
}
\end{equation}
Here, $\qsub$ is determined by the subshock compression ratio
\iec
\begin{equation}
\label{eq:qsub}
\qsub = 3\Rsub /(\Rsub - 1)
\ ,
\end{equation}
$\qmin$ is given by,
\begin{equation}
\label{eq:qmin}
\qmin =
3.5 + \frac{3.5 - 0.5\, \Rsub}{2\, \Rtot - \Rsub - 1}
\ ,
\end{equation}
(see Berezhko \cite{Berez96} for a full discussion of this equation and
Malkov \cite{Malkov97}, \cite{Malkov99} for an alternative derivation) and
\begin{equation}
\label{eq:qint}
\qint = (\qTP^{\prime} + \qmin)/2
\ ,
\end{equation}
where the prime indicates that we use
\begin{equation}
\Rtot^{\prime} = { (\gamgas +1) \MSZ^2 \over
{(\gamgas -1) \MSZ^2 + 2}} \le 4
 \label{eq:Rtotprime}
\end{equation}
to calculate $\qTP^{\prime}$. The far upstream sonic Mach number is
$\MSZ$, $\gamgas = 5/3$ is the adiabatic index of the shocked gas
excluding cosmic rays, and the normalization factors are
$\amc = \ainj \, [\pinj/ ( \mpc)]^{\qsub}$,
and
$\amax = \amc \, (\mpc / \pinterm)^{\qint}$.
The full explanation of these terms (including the normalization
$\ainj$ in terms of input parameters) is given in Berezhko \& Ellison
(\cite{BEapj99}), but for here it is only necessary to state that the
nonlinear spectrum can be determined with four arbitrary parameters:
(1) $\pinj$, (2) the rate of proton injection, $\etainjP$, (3) the
maximum momentum, $\pmax$, where the spectrum cuts off, and (4)
$\pinterm$. The results are relatively insensitive to $\pinterm$ which we
take as $\pinterm = 0.01 \pmax$ if $\pinterm > m_p \, c$ or $\pinterm = m_p \,
c$ otherwise, and $\pinj$ and $\etainjP$ are related, leaving
essentially two free parameters to describe the proton spectrum (here
and elsewhere, $m_p$ is the proton rest mass).  In this steady-state
description, the proton injection {\it rate}, $\etainjP$
(Berezhko et al. use the notation $\eta$),
is equivalent to the fraction of
total shocked proton number density, $n_2$, in
protons with momentum $p \ge \pinj$.
The scheme we adopt for determining the scale length that fixes $\pmax$ is
described in Section~3.1.

The thermal portion of the downstream spectrum is taken to be a
Maxwell-Boltzmann distribution at the shocked temperature and density,
\iec
$T_2 = P_2 \, m_p /(\rho_2 \, k)$,
where the downstream pressure, $P_2$, and downstream density, $\rho_2
= \rho_0 \, \Rtot$, are determined by the solution,
\iec they couple to the shock dynamics.  No extra
parameters are required to specify the thermal portion of the proton
distribution other than to assume it is Maxwell-Boltzmann.
In Figure~\ref{fig:EBfive} (Fig.~5 from \BEname\ \cite{BEapj99}), we
compare results from the simple model with those of a \MC\ shock
simulation used as a reference. The \MC\ simulation calculates the
entire spectrum from thermal energies upward and has an internally
consistent determination of the injection efficiency. Using this value
of $\etainjP$ in the simple model (the solid triangle is at $\pinj$)
gives an excellent match between the two techniques.  The extremely
high compression ratios occur because only adiabatic heating of the
shock precursor is assumed. When \alf\ wave heating is included (as it
is in all of the results presented below except those shown in
Figure~\ref{fig:fpdjde}), considerably smaller $\Rtot$'s
result (see Berezhko \& Ellison \cite{BEapj99} for a discussion of
\alf\ wave heating).\footnote{Helium is included in the calculation of
Eqs.~(\ref{eq:ThreePower}) in an approximate way by adding a pressure
contribution equal to $n_{\mathrm He}/n_{\mathrm H}$ times the
energetic proton pressure.  This neglects any enhancement effects for
heavy ions, but these could be incorporated in the simple model with a
parameterization such as used by Berezhko
\& Ksenofontov  (\cite{BereKsen99}).
Based on the work of Baring \etal\
(\cite{BaringEtal99}), we expect the contribution of helium to the pion-decay
emission to be no more than 50\% 
of the proton contribution.}

\subsection{Electrons}
 \label{sec:simple_e}

\placefigure{fig:fpdjde}

Before describing how electrons are treated in our simple model, we
note that  electron injection is harder to characterize than proton
injection for several fundamental reasons.  One
stems from the basic point that the shock structure is
determined almost totally by the ions, since they carry most of the
momentum, so general considerations of momentum and energy
conservation constrain their behavior.  Internally-consistent models of
ion injection can be developed without a detailed knowledge of the
complex plasma processes, and these models have been shown to match
spacecraft observations of ion acceleration at heliospheric shocks
(\egc Ellison, M\"obius, \& Paschmann \cite{EMP90}; Baring \etal\
\cite{BaringEtal97}) and hybrid simulations (\egc Giacalone \etal\
\cite{GBSE97}).  Electrons, on the other hand, carry little momentum and don't
influence the shock structure substantially so they act basically as
test particles.  Because of this, their injection and acceleration
efficiencies (at nonrelativistic energies at least)
are sensitive to the details of the complex wave-particle plasma
interactions, which are not well understood. Furthermore, since
protons determine the overall shock dynamics, plasma simulations
wishing to describe electrons self-consistently must include protons
self-consistently as well, forcing prohibitively large ranges in time
and length scales.
Compounding the problem, spacecraft observations of heliospheric
shocks have not until very recently (\iec Terasawa \etal\
\cite{Terasawa99})
detected the injection and acceleration of {\it thermal}
electrons. This, in contrast to the many observations of thermal
proton injection and acceleration, has made it impossible to infer the
properties of electron injection and acceleration from in situ
observations, or even, until recently, confirm that diffusive shock
acceleration of electrons from the thermal background takes place.

In order to include electrons in the simple model we note that
particles with the same upstream diffusion length, $\kappa/u_0$
($\kappa$ is the diffusion coefficient and $u_0$ is the far upstream
shock speed),\footnote{Everywhere, the subscript `0' implies far
upstream values and the subscript `2' implies far downstream values.}
have the same acceleration rate. That is, electrons and protons will
obtain the same spectral shape if they have the same diffusion
coefficient $\kappa_e(p)=\kappa_p(p)$ in a considered momentum
range. Since the diffusion coefficient $\kappa = \lambda(R)
\, v/3$ depends not only on the particle rigidity, $R$, but also on
their speed, $v$, which is different for electrons and protons at
$p<$1~GeV$/c$, one might conclude that, in general, the shapes of the
accelerated electron and proton spectra are different.  However,
it is easy to show that the electron and proton spectral shapes
should, in fact,
be very similar at superthermal energies.  At relativistic
energies, $E \gg m_p \, c^2$, the interaction with the magnetic field
of electrons and protons of the same gyroradius (or rigidity) is
indistinguishable if we assume any effects from helicity are
insignificant.  Therefore, the shape of the electron spectrum,
$f_e(p)$, at $p \gg m_p\, c$ is the same as the proton spectrum,
$f_p(p)$. Furthermore, to the accuracy of the simple proton model,
where we assume that protons with $p\ge m_p\, c$ have speed $c$,
and those with $p < m_p\, c$ have speed $v =p/m_p$,
the spectral shapes will be the same for all $p \ge m_p\, c$.
To the same accuracy, electrons with momenta $p < m_p\, c$ have
smaller diffusion lengths than protons at $p=m_p \, c$ and they `feel'
only part of the shock transition, i.e., the subshock as assumed in
the simple model. Therefore, just as we assume for the protons, the
spectrum of accelerated electrons at $p<m_pc\,$ can be approximated by
the power law $f_e\propto p^{-\qsub}$, independent of the diffusion
coefficient $\kappa_e(p)$, as long as $\kappa$ is an increasing
function of momentum.

Thus within the accuracy of the simple model,
\begin{equation}
\label{eq:ElecProton}
f_e(p) =
K_{ep} \, f_p(p),
\end{equation}
at all superthermal energies,
where the value of the numerical factor, $K_{ep}$, is determined
by the relation between the electron and proton injection rates,
i.e., using Eq.~(\ref{eq:ThreePower}) we have
$K_{ep}=(\ainjE /\ainjP)\, [\pinje /\pinjp ]^{\qsub}$, where the
subscripts $e$ and $p$ denote electrons and protons, respectively.
Unfortunately there are no reliable theoretical predictions for this
relation so we parameterize it.
We remark that if the damping of high frequency waves near thermal
energies is strong enough to inhibit electron scattering, there will
exist a range of suprathermal electron momenta where the electron
diffusion coefficient and length are larger than the proton ones, \iec
$\kappa_{\mathrm{e}} > \kappa_{\mathrm{p}}$ (\egc Levinson
\cite{Levinson92}, \cite{Levinson94}).  This case is considered
in Baring \etal\ (\cite{BaringEtal99}), where the {\it in}efficient
scattering of electrons causes their diffusion lengths to lengthen and
{\it increases} their steady-state injection efficiencies.  If this is the
case, the electron spectra in the sub-GeV range may be quite different
from those shown here and these differences can be particularly
relevant to radio synchrotron and nonthermal \brem\ emission.

If we make the specific assumption that the scattering mean free path
is 
\begin{equation}
\label{eq:mfp}
\lambda = 
\etamfp \, \rgmax \, \left ( \rg / \rgmax\right )^\alpha
\ ,
\end{equation}
where 
$\etamfp$ is a
parameter independent of particle momentum (Baring \etal\
\cite{BaringEtal99}, use the notation $\eta$), $\rg= p/(q B)$ is
the gyroradius in SI units, $\rgmax$ is the gyroradius at the maximum
momentum, $\pmax$, and $\alpha$ is a constant parameter ($\alpha = 1 =
\etamfp$ is roughly the Bohm limit), then electrons and protons which
satisfy,
\begin{equation}
\label{eq:momspec}
\etamfpElec \, p_{\mathrm{e}}^{\alpha} \, v_{\mathrm{e}} =
\etamfpPro \, p_{\mathrm{p}}^{\alpha} \, v_{\mathrm{p}}
\ ,
\end{equation}
obtain the same spectral shape. Since we obtain the proton spectrum
from the simple model, and if we assume that $\etamfp$ is the same for
electrons and protons, Eq.~(\ref{eq:momspec}) allows us to determine the
electron spectrum by finding the momenta dividing the three-component
power law, \iec the intervals where $\qsub$, $\qint$, and $\qmin$ in
Eq.~(\ref{eq:ThreePower}) apply.
We note that the above is just for concreteness and that, within the
approximations of our simple model, Eq.~(\ref{eq:ElecProton}) holds
and we do not have to make any
specific assumptions concerning the form of the particle diffusion
coefficient other than that it is a monotonically
increasing function of momentum.

In the top panel of Figure~\ref{fig:fpdjde} we show examples of
$f_e(p)$ (dashed and dotted curves) and $f_p(p)$ (solid curve)
obtained assuming Eq.~(\ref{eq:mfp}) with $\alpha=1$. As just
explained, electrons and protons have identical superthermal shapes
apart from a slight offset near $m_p \, c$ caused by assuming
Eq.~(\ref{eq:momspec}) rather than Eq.~(\ref{eq:ElecProton}).

To set the amplitude of the superthermal electron spectrum, we find it
convenient to use the electron to proton density ratio at relativistic
energies, $\epRatio$.  We chose $\epRatio$, rather than some injection
parameter, $\etainjE$, analogous to $\etainjP$, since $\epRatio$ can
be constrained by observations such as those of Galactic cosmic rays,
which suggest values around 1--5\% (\egc M\"uller \& Tang
\cite{Tang87}; M\"uller \etal\ \cite{Mueller95}) in the 1--10 GeV
range.  The final parameter needed to completely specify the electron
spectrum is the shocked electron to proton temperature ratio,
$\TempRatio$,
for which there is considerable latitude given uncertainties in the
shock speed and other parameters for individual SNRs that leaves
$T_{\rm p2}$ ill-determined observationally.  Note that detections of
X-ray emission (\egc Zimmermann, Tr\"umper, \& Yorke
\cite{Zimm96}) 
can constrain $T_{\rm e2}$ if contributions from nonthermal electron
components can be ruled out.  Given $\TempRatio$ and $\epRatio$, it is
straightforward to compute $\etainjE$, as follows.  The number density
of particles per unit energy interval, $dN/dE$, is given by,
\begin{equation}
\frac{dN}{dE} =
4 \pi p^2 f(p) \, \frac{dp}{dE}
\ ,
\end{equation}
where $dp/dE = (E + m_0\, c^2) / (p\, c^2)$, for $m_0=m_e, m_p$ as the
case may be.  Here and elsewhere, $E$ is the kinetic energy.
Exploiting the fact that the electron and proton phase space
distributions have essentially the same shape we can write the ratio at
a given  nonrelativistic energy, $E_e=E_p=\EnNR \ll m_e\, c^2$,
\begin{eqnarray}
\frac{\etainjE}{\etainjP}
\equiv
\frac{\left( dN/dE \right)_e}
     {\left( dN/dE \right)_p} \Biggr\arrowvert_{\EnNR}
=
\frac{\ainjE \, m_e \, p_e \, \left ( p_e/\pinje \right
)^{-\qsub}}
     {\ainjP \, m_p \, p_p \, \left ( p_p / \pinjp \right )^{-\qsub}}
=
\frac{\ainjE}{\ainjP} \,
\left ( \frac{m_e}{m_p} \right )^{3/2} \,
\left ( \frac{T_{\rm e2}}{T_{\rm p2}}\right )^{\qsub/2}
\ ,
 \label{eq:NRRatio}
\end{eqnarray}
since $\pinje\propto (m_eT_{\rm e2})^{1/2}$, $p_e\propto (m_eE_e)^{1/2}$
and likewise for protons.  This can be inverted to yield
$\ainjE/\ainjP$.
At identical relativistic energies, $E_e=E_p=\Enrel \gg m_p\,
c^2$, since $\pinterm$ is identical for protons and electrons, we have,
\begin{equation}
\label{eq:RelRatio}
\epRatio
\equiv
\frac{\left( dN/dE \right)_e}
     {\left( dN/dE \right)_p} \Biggr\arrowvert_{\Enrel}
=
\frac{p_e^2 \, \amaxe \, \left ( p_p / \pinterm \right )^{-\qmin} }
{p_p^2 \, \amaxp \, \left ( p_e / \pinterm \right )^{-\qmin} }
=
\frac{\amaxe}{\amaxp} =
\frac{\ainjE}{\ainjP} \left ( \frac{m_e}{m_p} \;
                 \frac{T_{\rm e2}}{T_{\rm p2}} \right )^{\qsub/2}
\ ,
\end{equation}
since $p_p = p_e$ at relativistic energies.
The last equality is obtained using the definition of $\amax$ given
just after Eq.~(\ref{eq:Rtotprime}).
Eqs.~(\ref{eq:NRRatio}) and~(\ref{eq:RelRatio})
define the relationship between $\etainjE$, $\TempRatio$,
and $\epRatio$, with the explicit dependence on $\TempRatio$
disappearing when $\ainjE$ is specified.

Comparison of Eqs.~(\ref{eq:NRRatio}) and (\ref{eq:RelRatio}) shows
that the number density ratio decreases by a factor of $(m_e /
m_p)^{(\qsub-3)/2}$ in going from fully nonrelativistic to fully
relativistic energies. For strong shocks, $\Rsub \sim 3$ and $\qsub
\sim 4.5$, so producing an electron to proton ratio of $\epRatio \sim
0.01$ at relativistic energies, as observed for galactic CRs, implies
that the electron injection rate at nonrelativistic energies must be
approximately equal to the proton rate, \iec $\etainjE \simeq
\etainjP$.  This conclusion holds whether the
thermal electrons and protons are in equipartition or not.  However,
it is highly likely that the injection energy scales strongly with the
shocked temperature, implying that $T_{e2} \sim T_{p2}$, \iec it is
unlikely that $\epRatio \sim 0.01$ can be obtained unless the shocked
electron and proton temperatures are nearly equal,\footnote{The work
presented here implicitly assumes that all of the accelerated protons
and electrons originate from the thermal background. In fact,
superthermal seed populations exist, such as the Galactic cosmic rays
and, in young SNRs, positrons and electrons from radioactive decay of
freshly synthesized material will be accelerated if they encounter
local shocks. For simplicity, however, we assume that the contributions
from these sources to the freshly accelerated thermal material is
small, 
noting that the ambient ISM density exceeds that of cosmic rays
by several orders of magnitude.}
a result suggested by some thermal X-ray signatures of SNRs (\egc
Zimmermann, Tr\"umper, \& Yorke \cite{Zimm96}).

For the example shown in Figure~\ref{fig:fpdjde}, we have chosen
$\TempRatio = 1$,
and used $\epRatio= 0.05$ (dashed line) and
$\epRatio = 0.01$ (dotted line).  The spectra also include an
exponential cutoff (not included in the initial \BEname\
\cite{BEapj99} work) so that $f(p)$ given by
Eqs.~(\ref{eq:ThreePower}) becomes (see Berezhko \& Krymsky
\cite{BereKrym88} and Berezhko \etal\ \cite{BerezEtal96}):
\begin{equation}
\label{eq:expo}
f(p) \rightarrow
f(p) \,
\exp{\left [ -\frac{1}{\alpha} 
\left ( \frac{p}{\pmax} \right )^{\alpha}
\right ]}
\ ,
\end{equation}
where the $\alpha$ is the same as in Eqn.~(\ref{eq:mfp}).  

The bottom panel of Figure~\ref{fig:fpdjde} shows the differential
flux, $dJ/dE = [v/(4 \pi)] dN/dE$, in particles per cm$^{2}$ per sec
per ster per MeV.  For $\epRatio = 0.05$, $\etainjE = 0.012$, while
for $\epRatio=0.01$, $\etainjE=2.5\xx{-3}$, if we assume that the
electron and proton injection energies are the same.

Finally, we define the acceleration efficiency, $\EffRel$, as the
fraction of total incoming energy flux that goes into
particles with momentum $p \ge \mpc$, \iec
\begin{equation}
\label{eq:EnEff}
\EffRel =
\frac{ [\gamma_2/(\gamma_2-1)] \, 
P_{\mathrm{rel,2}}(> \mpc) \, u_2 + F_E }{(1/2)
\rho_0 u_0^3 + (5/2) \, P_0 u_0}
\ ,
\end{equation}
where $P_{\mathrm{rel,2}}(>\mpc)$ is the downstream pressure in
particles of momentum $\mpc$ or greater, $u$ is the bulk flow speed in
the shock frame, $\gamma_2$ is the effective downstream ratio of
specific heats including the effects of relativistic particle
pressure, and we have taken the ratio of specific heats for the
unshocked plasma to be $5/3$.  The escaping energy flux, $F_E$, is
included in this definition since these are accelerated particles that
only leave the system after obtaining an energy $E \sim \Emax$. The
efficiency for selected examples are listed in Tables~1 and 2 under
`Output values'.

\placefigure{fig:photonOne}            

\subsection{Photons}
 \label{sec:simple_phot}

The photon emission processes considered in this paper are those
discussed at length in Baring \etal\ (\cite{BaringEtal99}): synchrotron
emission that pertains to the radio to X-ray wavebands, inverse
Compton scattering of cosmic microwave background radiation that
contributes to the gamma-ray signal, X-ray/gamma-ray bremsstrahlung
emission of thermal and energetic electrons interacting with both
ambient electrons and protons,\footnote{We note that,
to correct some typographical errors and improve the accuracy of an
approximation to the $e$--$e$ bremsstrahlung cross-section, Eqs.~(A2)
and (A4) in Baring \etal\ (\cite{BaringEtal99}) should be modified
such that all $\gamma_e$'s in Eq.~(A2) are replaced by $\gamma_e - 1$,
and Eq.~(A4) there is written as
\begin{equation}
A(\engam,\,\entot) =
1 - \frac{10}{3}\,
{ (\entot - 1)^{1/5} \over {\entot + 1}}
\left ( {\engam \over {\entot}} \right )^{1/3}
\end{equation}
for $\engam$ where $A(\engam,\,\entot)$ is positive; otherwise
$A(\engam,\,\entot)$ is set to zero.}
and pion decay gamma-rays spawned by
$p$--$p$ collisions involving the accelerated cosmic rays.
As in Baring \etal\ (\cite{BaringEtal99}), we explicitly include 
the photon production from He--$p$ and He--He collisions.
While Sturner \etal\
(\cite{SturnerEtal97}) were the first to examine the radiative
signatures of all of these components, Baring \etal\
(\cite{BaringEtal99}) were the first to treat them all in the context
of a detailed model of nonlinear shock acceleration.  Generally,
infrared radiation fields peculiar to individual remnants contribute
only a minority of the inverse Compton signal (\egc Gaisser,
Protheroe, \& Stanev \cite{GPS97}; but see de Jager \& Mastichiadis'
\cite{deJager97} discussion of the environment of W44).  To date, and
also in this paper, synchrotron self-Compton contributions are
neglected, being important only in strongly magnetized (\iec
$B\gtrsim 100\mu$G) remnants such as Cas A.  Furthermore, inverse
bremsstrahlung from fast ions contributes insignificantly in general
(Baring, Jones, \& Ellison \cite{BJE2000}), and is accordingly neglected
here.

Losses from \syn\ and \IC\ emission are included and combined
such that electron spectra cutoff at an energy
\begin{eqnarray}
\label{eq:synloss}
\Emax \simeq
\lefteqn{1.6\xx{5} \,
\etamfp^{-1/2} \,
\left ( 1 - \frac{1}{\Rtot}\right)^{1/2}\,
\left ( \frac{\Vsk}{\rm m/s}\right ) \times } \nonumber \\
& & 
\left [ \left ( \frac{B_0}{\rm{G}}\right )^{-1} + \Rtot
  \left ( \frac{B_2}{\rm{G}} \right )^{-1} \right ]^{-1/2} \,
\left [ \left ( \frac{B_2}{\rm{G}}\right )^{2} +
\left ( \frac{\Bcbr}{\rm{G}}\right )^{2}
\right ]^{-1/2}
\rm{eV}
\ ,
\end{eqnarray}
where $\Vsk$ is the shock speed and $\Bcbr = 3.23$ \muG\ is the
equivalent magnetic field of the cosmic background radiation and
accounts for \IC\ losses.  
This expression differs slightly from that in Baring \etal\ because
we explicitly allow for the downstream field $B_2$ to be
different from $B_0$. Also, in order to be
consistent with the work of
\BKP (\cite{BereKP99a}),
we take the perpendicular component of the downstream magnetic field
to the line of sight to be $B_{2\perp} = 0.5 B_2$ in calculating the
\syn\ emission.

\section{MODELING SNR PHOTON EMISSION}
 \label{sec:SNR_emission}

Instead of arbitrarily choosing the shock speed and maximum
acceleration energy, we use a model of SNR evolution by Truelove \&
McKee (\cite{Truelove99}) that continuously maps between the free
expansion and Sedov phases to give the shock parameters as a
function of explosion energy, $\EnSN$, ejecta mass, $\Mej$, and
remnant age, $\tSNR$.
For this work, we only consider uniform distributions for the ambient
ISM mass,
appropriate to Type Ia progenitors. We 
do allow for a power-law density profile for the ejecta, as in
Chevalier (\cite{Chev82}), \iec
\begin{equation}
\label{eq:ejectapower}
\rho \propto
\tSNR^{-3} \, (r / \tSNR)^{-n}
\ ,
\end{equation}
where $\rho$ is density, $r$ is radius, and $n$ is a constant, but
only consider the forward shock. Examples with forward and reverse
shocks are given in Ellison (\cite{ellison2000}).
We caution that care must be used in applying
our acceleration model and associated predictions of
broad-band emission to Type II supernov\ae, which may 
explode in environments that are far from uniform, being strongly
modified by the stellar wind generated by their massive progenitors.

In Figure~\ref{fig:SNRcomp}a
we show the {\it instantaneous}
particle spectra produced by the forward shocks in two examples which
differ widely in ambient parameters.
We have a high density, high ejecta mass, high $B$-field   
example (`H' with heavy-weight curves) with proton number density
$\ProDenUpS = 10$ \pcc,
$\Mej = 5 M_{\sun}$, and
$B_0 = 30\xx{-6}$ G (more typical of Type II SN\ae),
and a low density, low ejecta mass, low $B$-field example (`L' with
light-weight curves) with
$\ProDenUpS = 0.01$ \pcc,
$\Mej = 1 M_{\sun}$, and
$B_0 = 3\xx{-6}$ G (more typical of Type Ia SN\ae).
A full listing of the parameters is given in Table~1 under columns
marked `High' and `Low' but all other input parameters are the same
between these two models, namely:
$\EnSN = 10^{51}$ erg,
$\tSNR= 1000$ yr,
$\TempRatio = 1$,
$\epRatio = 0.03$, and
$\etainjP = 1\xx{-3}$.
Figure~\ref{fig:SNRcomp}b shows photon fluxes at the Earth assuming a
distance, $\dSNR = 1$ kpc and an emission volume, $\EmisVol = 1$
pc$^3$. All output parameters are given in Table~1.  In all of the
examples in this paper we assume the plasma is fully ionized and the 
ambient helium number density to
be $0.1 \, \ProDenUpS$. We further assume that the unshocked helium
temperature {\it per nucleon} equals the proton temperature,
$\TempPro$ (this is typically observed in the heliosphere), and that
the unshocked electron temperature equals $\TempPro$.

There are some striking differences between the photon spectra for
these parameter sets which represent a reasonable spread for
various types of SNRs.  First, the three order of magnitude difference
in $\ProDenUpS$ and one order difference in $B_0$ produce much stronger
emission, per unit emission volume, for the high values.  Next, for
high values, the TeV emission is dominated by \pion\ with
\IC\ being completely insignificant (and not shown in
Figure~\ref{fig:SNRcomp}b), while for low values, the TeV emission is
dominated by \IC. The \IC\ from the low value example also yields the
highest photon energies.
At hard X-ray energies (10-100 keV), the emission from the high
density example has a strong nonthermal \brem\ component, while for the low
density example, the X-rays $< 100$ keV are almost
totally from \syn\ emission
from high energy electrons.
Deducing the shifting dominance of the various emission processes as
parameters vary is an important goal of this work.

\subsection{Parameter Survey}
 \label{sec:parameters}

\placefigure{fig:paraModel}            

\newlist

As indicated by the previous examples, nonlinear shock acceleration is
complicated and the solutions we describe have a large number of
parameters, even in the homogeneous environment we assume.
The input parameters are:
$\EnSN$, $\Mej$,
$\tSNR$,
$\ProDenUpS$,
$B_0$, $n$,
$\dSNR$, and
$\EmisVol$.
\newlist
Added to these are the parameters intrinsic to the model:
$\etamfp$,
$\etainjP$,
$\TempRatio$, and
$\epRatio$,
plus what we call `general model' assumptions. These include
$\alpha$, $\pinterm$, equation~(\ref{eq:expo}), and the assumptions of
ambient and ejecta mass distributions.  The maximum momenta
where the spectra cut off are also model parameters, but they are
determined from the SNR model ($\etamfp$ becomes important here)
as was described in Baring
\etal\ (\cite{BaringEtal99}).  That is, in determining the maximum
momenta of protons, we first assume equation~(\ref{eq:mfp}) and then
set the diffusive shock acceleration time, $\tacc$, equal to $\tSNR$,
or set the upstream diffusion length equal to $1/4$ of the shock
radius, whichever produces the lowest $\pmax$.\footnote{In determining
$\pmax$ at time $\tSNR$, we make the approximation that the shock speed
is constant at the value it has at $\tSNR$. In fact, an accurate
determination of $\pmax$ in an evolving remnant requires a more
complete model than we use here, i.e., Berezhko (\cite{Berez96}), which
keeps track of the history of particles, adiabatic losses, and the
numbers of particles accelerated at a given epoch. Despite this, our
values will not differ by large factors from more realistic ones.}
For electrons, $\pmax$ is equal to that of the protons or to the value
determined from combined \syn\ and \IC\ losses (see Baring \etal\
\cite{BaringEtal99}, for details), whichever is less.

\subsubsection{Emission volume}
 \label{sec:volume}

For a uniform environment, we can estimate the emission volume in the
following way.  The contact discontinuity separates the swept up ISM
material from the ejecta and acts as the piston for the forward
shock. During the free expansion and early Sedov phases, the swept up
material forms a nearly uniform, dense shell behind the forward shock
such that,
\begin{equation}
\label{eq:shell}
\Rsk^3 \, \rho_0 \approx
\rhoShell \, (\Rsk^3 - \Rpiston^3)
\ ,
\end{equation}
where $\Rpiston$ is the radius of the contact discontinuity, $\rho_0$
is the density of the background ISM, and $\rhoShell$ is the density
in the shell. In the approximation that the shell has uniform density,
$\rhoShell = \Rtot \, \rho_0$, \iec it has the shocked density,
$\rho_2$.
For large compression ratios, the thickness of the shell, $L = \Rsk -
\Rpiston \ll \Rsk$, and we can write,
\begin{equation}
\frac{1}{\Rtot} =
1 - \left ( \frac{\Rpiston}{\Rsk} \right )^3 =
1 - \left ( \frac{\Rsk - L}{\Rsk} \right )^3 \approx
\frac{3 \, L}{\Rsk}
\ .
\end{equation}
During later phases, the contact discontinuity will be far from the
forward shock but we can then interpret $L$ as the thickness of the
dense shell behind the forward shock where most of the emission occurs
independent of the position of the contact discontinuity.
With these approximations, the emission volume can be estimated as
\begin{equation}
\label{eq:Emission}
\EmisVol \approx (4\pi/3) \,\Rsk^3 / \Rtot
\ ,
\end{equation}
so that the emission volume generally is considerably less than the
total remnant volume.

\subsubsection{Trends and Key Parameters}
 \label{sec:keyparameters}

\placefigure{fig:paraModel}            
In Figures~\ref{fig:paraModel}, \ref{fig:paraInputSens}, and
\ref{fig:paraInputBot} 
we plot the total photon emission varying a single parameter as
indicated.  All of these results are normalized to $\dSNR = 1$ kpc\
and $\EmisVol$ as given in Eqn.~(\ref{eq:Emission}).  Varying
$\etainjP$, $\TempRatio$, or $\epRatio$ (Fig~\ref{fig:paraModel})
produces relatively modest changes in the photon emission,
particularly in the radio to X-ray and gamma-ray bands.  An exception
occurs for $\etainjP=10^{-5}$ (solid curve,
Fig.~\ref{fig:paraModel}a).  This injection rate is low enough that a
{\it test-particle} result occurs even though the Mach number is quite
high ($\MSZ \simeq 130$), i.e., the energy placed in superthermal
particles is not enough to smooth the shock.  A test-particle solution
is different in two main ways from the nonlinear cases: the overall
compression ratio, $\Rtot$, is approximately four which is
considerably less than for the higher injection rate cases, producing
the steeper spectrum at the highest energies, and the shocked
temperature is considerably higher than the nonlinear temperatures (as
indicated by the position of the UV/X-ray thermal bremsstrahlung peak
in panel \ref{fig:paraModel}a).  While the gamma-ray flux drops in the
test-particle regime, the concurrence of curves for higher $\etainjP$
underlines the regulatory effect (\egc Eichler \cite{Eich84}; Ellison
\& Eichler \cite{EE84}; see also the review of Jones and Ellison
\cite{JE91}) of nonlinear acceleration.

Fugure~\ref{fig:paraModel}b reveals that by itself, the value of the
parameter $\TempRatio$ is of little significance.  This is because of
its coupling to $\etainjP$, and $\epRatio$, as described in
Section~\ref{sec:simple_e}, so that by fixing $\etainjP$, and
$\epRatio$, the non-thermal X-ray/soft gamma-ray bremsstrahlung flux
and the hard gamma-ray pion decay and inverse Compton contributions
remain approximately fixed.  
The proton temperature is determined by the shock dynamics
solution.

Changes in the electron to proton ratio at relativistic energies,
$\epRatio$ (Fig.~\ref{fig:paraModel}c), also produce important effects
in the superthermal
\brem\ range,  since 
this component depends
on the normalization of the non-thermal electron
distribution.  
As $\epRatio$ increases, the 
super-GeV gamma-ray band shifts from being \pion\ dominated to 
having nearly equal \brem, \IC, and \pion\ components. 
We show below examples with a more
tenuous ISM where \IC\ dominates
the gamma-rays so that the flux in this
energy range becomes quite sensitive to the choice of $\epRatio$.

\placefigure{fig:paraInputSens}            
\placefigure{fig:paraInputBot}            

The ambient density is expected to have a large natural range
(Fig.~\ref{fig:paraInputSens}a) and strongly influences 
the emitted
flux. Low values of $\ProDenUpS$ give the highest
maximum energy with the $>$TeV flux dominated by \IC. 
As $\ProDenUpS$ increases, the
\pion\ component becomes more pronounced in the 100 MeV to TeV range
and \brem\ becomes dominant in the X-ray band. 
X-ray emission
is dominated by \syn\ at low $\ProDenUpS$, and the radio emission is
relatively insensitive to $\ProDenUpS$ 
with a slight flattening occurring at higher densities.
Naively, one expects that the synchrotron and inverse
Compton components should scale linearly with density increases, while
the bremsstrahlung and pion decay contributions should be proportional
to $\ProDenUpS^2$.  However, the global spectral properties are an
accumulation of effects caused by the complexity of the nonlinear
acceleration mechanism and the evolution of the SNR as given by the
Truelove \& McKee parameterization.  
As $\ProDenUpS$ rises, the expanding supernova sweeps up its ejecta
mass sooner, and therefore decelerates on shorter timescales, thereby
reducing both the radius and volume of a remnant of given age, and
lowering the shock speed, the compression ratio,  and the
downstream proton temperature $T_{\rm p2}$ [\egc see Eq.~(7) of Baring
\etal\ \cite{BaringEtal99}; and Table~2].  
The bottom line is that the normalization of the overall distribution
and the relative contributions of the various components, are not
simple functions of density and can vary strongly in narrow energy
bands.
The full set of
parameters for Figure~\ref{fig:paraInputSens} is given in Table~2.

Some remarkable features in Figure~\ref{fig:paraInputSens}a (where
$B=3\mu$G for all four examples) include 
(i) for $\ProDenUpS > 1$
cm$^{-3}$ (\iec the dotted curve, see
also Figure~\ref{fig:SNRcomp}), the radio to optical/UV emission is of
thermal bremsstrahlung rather than synchrotron origin, 
(ii) the highest energy emission occurs for the lowest $\ProDenUpS$
and is dominated by \IC,
and (iii) the spectral shape in the X-ray band (1-100 keV)
is highly sensitive to
the density (thereby providing powerful observational diagnostics) and
can have a strong nonthermal \brem\ component at high $\ProDenUpS$.  

Variations caused by adjusting the ambient
magnetic field are illustrated in Fig.~\ref{fig:paraInputSens}b (where
$\ProDenUpS=1$ cm$^{-3}$ for all four examples).  The principal property
immediately apparent is the anti-correlation between radio and GeV-TeV
fluxes.
The higher the value of $B_0$, the brighter the
radio synchrotron 
[scaling as $\sim B^{3/2}$; \egc see Eq. (6.36) of Rybicki
\& Lightman \cite{RL79}], 
which is the effect usually incorporated in test-particle
SNR models (e.g. Mastichiadis \& de Jager \cite{MdeJ96}). 
In addition, however,
the hard gamma-ray emission becomes fainter with increasing $B_0$.
This property arises mostly because of the influence of the magnetic
field on the shock dynamics and total compression ratio.  
The Alfv\'enic Mach number drops and \alf\ wave heating in the
precursor becomes stronger as $B_0$ increases causing $\Rtot$ to
decrease.  This weakening of the shock steepens the particle
distributions and the overall photon spectrum, with the net effect of
a reduction of the gamma-ray flux together with a change in its
spectral index.\footnote{The spectrum is further steepened because the
\alf\ waves which do the scattering are assumed to move through the
upstream plasma such that they move with the \alf\ speed away from the
shock when viewed from the local plasma frame.}
The emission from radio through the X-ray band is dominated by \syn\
emission from energetic electrons for $B_0 > 10$ \muG, with \brem\
dominating X-rays for $B_0 \le 3$ \muG.  Another consequence of a rise
in $B_0$ is that the maximum {\it ion} energy increases while the
maximum {\it electron} energy declines due to \syn\ losses.  This
causes \pion\ to dominate the gamma-ray band and
the UV/soft X-ray synchrotron peak becomes approximately independent
of $B_0$, a well-known property (\egc de Jager
\etal\ \cite{deJagerEtal96}; 
Reynolds \cite{Reynolds96}) of cooling-limited synchrotron radiation.
Note again that we omit the regime $B_0>100\mu$G where the Alfv\'enic
Mach number is as low as a few and where second-order Fermi
acceleration is important (see Bykov
\etal\ \cite{Bykov2000} and references therein), an effect 
beyond the scope of the present work.

The plots in Fig.~\ref{fig:paraInputSens} include the
approximate integral flux sensitivities (adjusted for the
$E_{\gamma}^2$ times differential flux representation) of the proposed
Veritas experiment (Weekes \etal\ \cite{weekes99}) and the planned
GLAST mission (Gehrels \& Michelson \cite{Gehrels99}).  
These represent the
probability that each of these experiments will detect photons above a
given energy, and clearly indicate that GLAST and Veritas will make
significant observational progress in the 100 MeV--100 GeV and 100
GeV--10 TeV bands, if SNRs emit according to the predictions here.
CELESTE provides somewhat better sensitivity in the limited 50-70 GeV
range, but we chose the Veritas sensitivity as a benchmark for
atmospheric
\v{C}erenkov telescopes (ACTs).  Several new TeV programs are in the
works, including MAGIC (Lorenz \cite{Lorenz97}) in the Canary islands,
and HESS (e.g. Kohnle \etal\ \cite{kohnle99}) and CANGAROO-III in the
southern hemisphere; their sensitivities differ only modestly in
energy range and flux from the ones depicted for Veritas.  Because of
the ``saturation'' of gamma-ray spectra at low densities, due to the
insensitivity of inverse Compton emission to ISM density, for
reasonable ISM fields these experiments should have no difficulty
detecting remnants throughout the Sedov epoch and prior to the
radiative phase.  The only way this cannot materialize is for shocks
at remnant shells not to be efficiently accelerating particles to
energies beyond around 10 GeV,
\iec contrary to general expectations.  
Therefore, we anticipate positive detections in a number of SNRs in
and out of the galactic plane, and even the most pessimistic case of
no detections will lead to an improved understanding
of particle acceleration.

In particular, remnants in moderately dense environs (i.e. $\gtrsim 1$
cm$^{-3}$) will be ideal candidates for GLAST to search for the
$\pi^0$ bump emission so clearly evident in the higher density cases
in Fig.~\ref{fig:paraInputSens}a.  A clearer indication of this can be
found in Baring (\cite{Baring00}), where these model spectra are
compared with the simulated {\it differential} flux sensitivity
expected for GLAST, a truer indicator of the ability of the instrument
to perform spectral measurements; Baring (\cite{Baring00}) observes
that the $\pi^0$ bump will be a clearly discernible spectral structure
when ISM densities exceed around 0.3~cm$^{-3}$.  Confirmation of the
existence of accelerated ions that spawn such a gamma-ray signature is
an unambiguous signature of cosmic ray production.  We note that the
EGRET sensitivity over the 100 MeV -- 10 GeV band is in the
$10^{-5}$--$10^{-4}$ MeV cm$^{-2}$ sec$^{-1}$ range, so that only
remnants with ambient densities exceeding around 0.3~cm$^{-3}$ would be
possibly detectable by EGRET.  Moreover, EGRET would only be sensitive
to pion decay bumps when the environmental densities are quite large,
well above 10~cm$^{-3}$.

In Fig.~\ref{fig:paraInputBot} we show variations in
the SN explosion energy and the age of the
remnant.
Adjusting $\EnSN$ clearly produces
little change in the broad-band spectral shape, merely modifying the
normalization due to changes in the radius and volume of the
remnant.  On the other hand, at early times (\iec  $\tSNR\lesssim
300$yr), the SNR age has a strong effect on the flux,
mainly because the remnant volume is small prior to the Sedov phase.
During these early times, which consists mostly of the free expansion
phase, the shock is extremely nonlinear in nature, obtaining a very
large compression ratio ($\Rtot \simeq 25$ when $\tSNR=30$ yr and
$\Rtot\simeq 24$ when $\tSNR=100$ yr),\footnote{Note that high Mach
number {\it unmodified} shocks (\iec those with $\Rtot \simeq 4$) are
possible as described in general by Berezhko \& Ellison
(\cite{BEapj99}) and specifically for SN1006 by Berezhko, Ksenofontov,
\& Petukhov (\cite{BereKP99a}) if injection rates lower than that
assumed here are used.}
and the gamma-ray band is dominated by \pion.
The shock speed is high during this time and the
downstream temperature, while very much less than the R-H value
($T_2 \simeq 3\xx{7}$K for the nonlinear shock at $\tSNR=100$ yr versus
$T_2\simeq 1.4\xx{9}$K for the test-particle shock!), is
also high.
The transition into the Sedov phase at $\tSNR\sim 300$ yr brings about
a steady decline in $T_{\rm p2}$ along with a gamma-ray flux which is
almost independent of SNR age, a property present in the more complex
models of Dorfi (\cite{dorfi91}), Drury, Aharonian, \& V\"olk
(\cite{DAV94}), Baring
\etal (\cite{BaringEtal99}), and Berezhko \& V\"olk
(\cite{BerezVolk97}).  
The origin of this
gamma-ray insensitivity to $\tSNR$ prior to a remnant's radiative
phase is an approximate compensation (Baring \cite{Baring00}) 
between the volume
that scales as $\tSNR^{6/5}$ (radius $\propto\tSNR^{2/5}$) in the
Sedov phase, and the shock speed (and therefore also the square root
of $T_{\rm p2}$) that declines as $\tSNR^{-3/5}$ so that a roughly
$E^{-2}$ particle distribution function has a normalization that
scales as $T_{\rm p2}\propto \tSNR^{-6/5}$.  
The contrast between the free
expansion and Sedov phases is underlined by the change in shock
parameters: at $\tSNR=30$ yr (solid curve) $\Vsk = 1.4\xx{4}$ \kmps,
$\MSZ = 850$, $\MAZ=2.4\xx{3}$, and $\Rtot = 25$, while at $\tSNR=10^4$
yr (dotted curve), $\Vsk = 490$ \kmps, $\MSZ = 31$, $\MAZ=90$, and
$\Rtot = 8.5$.

\placefigure{fig:TeVtoRadio}            
\placefigure{fig:paraNormBot}            

In Figure~\ref{fig:TeVtoRadio} we show, as a function of $B_0$ and
$\ProDenUpS$, the following ratio:
\begin{equation}
\label{eq:ratio}
\Rphot =
\frac {\frac{1}{E_0}\,
\int_{E_0}^{\infty} \Fnu \, dE }
{\Fnu(E=10^{-11} {\rm MeV})}
\ ,
\end{equation}
that is, the integral photon flux above $E_0$ normalized to $E_0$ over
the radio flux ($\Fnu$ has units of cm$^{-2}$-s$^{-1}$ and we
set $E_0 = 500$ GeV in Fig.~\ref{fig:TeVtoRadio}).  
This ratio is critical for gamma-ray searches of radio SNRs and 
the most important result is that, for $\ProDenUpS \le 1$ \pcc, 
SNRs in high $B_0$ environments have the lowest $\Rphot$.
For $\ProDenUpS$ held fixed at 0.01 \pcc,
$\Rphot$  decreases by more than two orders of
magnitude as $B_0$ increases from 1 to 100 \muG.  However, the
absolute TeV flux depends strongly on the density as shown in
Figure~\ref{fig:paraInputSens}a.

Another critical factor for TeV observations
involves the relative importance of \IC\ versus \pion\ emission and the
impact this has on our ideas for the origin of cosmic rays. 
In Figure~\ref{fig:ic_pp} we show the \IC\ and
\pion\ components for the parameters marked with dots in
Figure~\ref{fig:TeVtoRadio}.  The variation in total absolute flux for
the three cases is relatively small but there is a shift from
\IC\ being dominant at TeV energies when $\ProDenUpS=0.01$ \pcc\ to \pion\
being dominant when $\ProDenUpS=10$ \pcc. In the high $B_0$ case, \pion\
emission extends to higher energies than the \IC\ because the electron
spectrum is cutoff from \syn\ and \IC\ losses.  Hence, atmospheric
\v{C}erenkov experiments may provide detections of pion decay emission
in sources with high magnetic fields.  This may be the case
in the recent marginal detection (V\"olk \etal\
\cite{Volk00}) of Cas A by HEGRA.  Yet the spectral 
shape in the TeV band bears no signature peculiar to \pion\ emission,
but rather is a marker of the underlying particle distributions.
Hence, the most powerful diagnostic the sensitive TeV experiments will
provide is the determination of the maximum energy of emission,
thereby constraining the ISM density, magnetic field and the electron
to proton ratio.

\placefigure{fig:paraNormBot}            

\subsection{SNR SN1006}
 \label{sec:SN1006_IC443}

\placefigure{fig:SN1006}         
\placefigure{fig:ic443}         

For a concrete example, we apply our model to SN1006,
the first remnant to have a detection of
shell-related TeV gamma-ray emission reported, by the CANGAROO
collaboration (Tanimori \etal\  \cite{Tanimori98}).
We constrain our parameters to match those used by \BKP\
(\cite{BereKP99a}) as closely as possible and discuss the comparison
with that model below. A full listing of the parameters is given in
Table~1 under the column labeled SN1006$_{\mathrm{NL}}$, where the `NL'
subscript contrasts this nonlinear result with the test-particle one
discussed in Section~\ref{sec:test-particle} below. The parameters
used by \BKP\ are listed in the column labeled BKP99.
In particular, we use an unshocked proton
density $\ProDenUpS = 0.1$ \pcc, SN energy $\EnSN =
10^{51}$ erg, ejecta mass $\Mej=1.4 M_{\sun}$ (\iec values similar to
those given by Truelove \& McKee \cite{Truelove99} and/or Laming
\etal\ \cite{LamingEtal96}), and an ejecta density profile
characterized by $n=7$ (\iec Eq.~\ref{eq:ejectapower}). 
For a remnant age $\tSNR = 990$ yr, these values yield $\Vsk \simeq
3750$ \kmps\ and $\Rsk \simeq 7$ pc in the Truelove and McKee model.
If we take the observed proper motion of $0''.30 \pm 0''.04$ yr$^{-1}$
(Long \etal\
\cite{Long88}), this speed yields a distance of 2.3 kpc, and using
the observed angular size of 30', this implies an angular radius of
10 pc, within 50\% of the 7 pc given above.

Figure~\ref{fig:SN1006} shows the continuum component photon
spectra obtained from these parameters, along with the sum of the components
(heavy solid line).
In obtaining the normalization for this fit, the emission volume was
taken to be $\sim 195$ pc$^3$, \iec the value given by
Eq.~(\ref{eq:Emission}) and the distance to the source was assumed to
be $\dSNR=1.7$ kpc, commensurate with that deduced from optical
observations (see Green \cite{Green98}, and references therein). If we
had taken $\dSNR=2.3$ kpc, the required emission volume would be
correspondingly larger to match the observed flux.  It's clear that a
reasonable fit to the observed spectra is produced considering that we
used the same parameters as \BKP.\footnote{We have used $\alpha=0.6$ for this fit.
\BKP\ (\cite{BereKP99a}) and Ammosov
\etal\ (\cite{Ammosov94}) found that a similar value ($\alpha=0.5$)
was required for a good fit and
Reynolds (\cite{Reynolds96}) gave a detailed explanation of why such a
broadening might occur for the electron spectrum.}
Note that we have not included free-free absorption in our models.

Our results indicate that the TeV emission in SN1006 is dominated by
the \IC\ component, in general agreement with the conclusions of
previous work that invoke test-particle power-laws, \iec Mastichiadis
\& de Jager (\cite{MdeJ96}) and Tanimori \etal\ (\cite{Tanimori98}).
However, there is a \pion\ component at TeV energies and small changes
in the parameters can make it more pronounced than shown in
Fig.~~\ref{fig:SN1006}, as indicated in the results of Berezhko,
Ksenofontov, \& Petukhov (\cite{BereKP99a},b) discussed below.
Also in agreement with previous work (\egc Koyama \etal\
\cite{Koyama95}; 
Reynolds \cite{Reynolds96}), the X-ray emission is dominated by \syn\
emission from super-TeV electrons.  However,
\brem\ does contribute some at hard X-ray energies and
again, this contribution is sensitive to the input parameters. The
result shown in Figure~\ref{fig:SN1006}
assumes $\TempRatio=1$.
AXAF and  XMM will probe the X-ray emission at
and below the ASCA data range depicted in Figure~\ref{fig:SN1006}, and
the INTEGRAL experiment will be sensitive to the hard X-ray band below
200 keV. Together, these spacecraft should be able to clearly
differentiate the \syn\ and \brem\ components in the X-ray range and,
therefore, provide constraints for the entire broad-band emission.

\subsubsection{Comparison with the Kinetic NL Model of Berezhko et
al.}
 \label{sec:BereCompare}

As mentioned above, Berezhko, Ksenofontov, \& Petukhov
(\cite{BereKP99a},\cite{BereKP99b}) have presented results
for SN1006 using a kinetic, nonlinear model of diffusive cosmic ray
acceleration by a spherically symmetric, expanding SNR.
This  model 
self-consistently solves the cosmic ray transport equations together
with the gas dynamic equations and is 
based on the work of Berezhko
\etal\ (\cite{BerezEtal96}) and Berezhko \& V\"olk
(\cite{BerezVolk97}).  It is more complete than the steady-state,
plane-shock model presented here and keeps track of the accelerated
particles as the SNR evolves and yields the integrated emission over
the entire remnant as a function of time including adiabatic
losses. It also considers the diffusive properties of the particles as
a function of energy so that low energy electrons responsible for the
radio emission occupy a thin region behind the forward shock, while
the highest energy particles fill the entire volume behind the shock.
Despite these substantial differences, both models produce good
fits to the broad-band continuum with virtually identical input
parameters.
The Berezhko, Ksenofontov, \& Petukhov parameters are
listed in Table~1 under the heading `BKP99.'

The only difference in the input parameters is the small difference in
$\alpha$, all other input parameters are the same.  Note that the
proton injection rate we use ($\etainjP = 2.1\xx{-4}$) is equivalent
to the value ($\etainjP=5\xx{-4}$) used by \BKP, since injection is
treated somewhat differently in the two models and these two values
give the same injection efficiency.
The output parameters, as indicated in Table~1, differ only slightly
and in Figure~\ref{fig:bereSN1006} we
compare the shapes of the proton distributions for the two models. The
most obvious difference is the lack of a thermal peak in the Berezhko
\etal\ result (solid curve). They inject particles at a superthermal
momentum (\iec at $\pinj$ as given in Eqn.~\ref{eq:ThreePower}) and do
not model thermal particles. However, they could have included a shock
heated thermal distribution as we do here, in which case it would have
been very similar to the dashed curve since both models have similar
pre-shock and injection conditions.  The good correspondence in shape
over the entire superthermal range occurs despite the fact that the
plane-shock, simple model is steady-state and uses the shock
conditions at the SNR age (as determined by the Truelove \& McKee
parameterization) to determine the emission at that instant. In
contrast, the Berezhko
\etal\ model is fully time-dependent and integrates the emission
over the remnant.  Notably, the spectral
curvature typical of nonlinear shock acceleration survives the
volume integration in the Berezhko \etal\
result.

The two models do differ somewhat in the fraction of TeV flux
contributed by \pion. \BKP\ find that \pion\ and \IC\ contribute about
equally while here, \IC\ is clearly dominant.  While these differences
and other effects will be considered with more detailed comparisons in
future work, we believe this example gives credence to our claim that
the simple model is getting the basic physics correct. Furthermore, the added
ability to provide rapid, broad-band fitting is very important because
it is simply not possible, because of computing
limitations, to perform a comprehensive parameter search with
complicated codes such as the Monte Carlo technique used in Baring
\etal\ (\cite{BaringEtal99}) in a timely manner.  
Our model does permit such a
search, and we anticipate it can be used to significant advantage in
astronomical data analysis packages, providing the capability, not
previously available, for incorporating the principal nonlinear
effects of shock acceleration.

\subsection{Test-particle vs. nonlinear results}
 \label{sec:test-particle}

As we mentioned above, despite considerable evidence to the contrary,
many astrophysical applications of diffusive shock acceleration still
assume that the acceleration is inefficient and produces test-particle power
laws.\footnote{Of course, many examples exist of full nonlinear
calculations of SNR evolution (\egc Dorfi \& V\"olk
\cite{DorfiV96}; Berezhko \etal\ \cite{BereKP99b}; Kang \& Jones
\cite{KJ95}), 
but despite this, we believe the importance of nonlinear effects has
been slow to reach the general astrophysical community.}  A high
efficiency for producing cosmic rays will have far-reaching
consequences for a number of applications because values for the
shocked temperature and density will differ from the test-particle
\RH\ values.  This will be particularly important for X-ray line
emission models.
To illustrate how the predictions of diffusive shock acceleration vary
depending on whether the test-particle or nonlinear assumption is
made, we show in Figure~\ref{fig:testpart} (light-weight curves) the
phase-space electron and proton distributions that produced the
emission shown in Figure~\ref{fig:SN1006}.  These are compared to
test-particle spectra (heavy-weight curves) calculated for parameters
which are the same except the proton injection efficiency is set to
$\etainjP = 1\xx{-5}$ in the test-particle case.

\placefigure{fig:testpart}         

For the test-particle case (column labeled SN1006$_{\mathrm{TP}}$ in
Table~1):
$\Rtot=4.1$, $\Rsub = 4.0$, and
$T_2 \simeq 1.7\xx{8}$ K, while for the nonlinear case:
$\Rtot=7.2$, $\Rsub=3.6$, and
$T_2 \simeq 4.7\xx{7}$ K.
In the nonlinear case, the shocked density is $\sim 75$\% {\it
greater} and the shocked temperature is more than a factor of 3
{\it less} than the test-particle result.
Furthermore, the nonlinear electron spectrum in
Figure~\ref{fig:testpart} shows a stronger nonthermal tail on the
thermal peak (comparison between the dotted and dot-dashed curves in
Figure~\ref{fig:fpdjde} gives a much more extreme example).  Accurate
electron temperatures and knowledge of the presence of significant
nonthermal electron distributions are of prime importance for modeling
X-ray line emission from shock heated gas (\egc Hamilton \etal\
\cite{HSC83}).  Line models depend strongly on the assumed electron
temperature and density and may change substantially if cosmic rays
are accelerated with efficiencies typical of the example just given.
Dorfi \& B\"ohringer (\cite{DorfiB93}) (see also Dorfi \cite{Dorfi94})
calculate the evolution of SNRs including cosmic ray production and
find examples where $\gtrsim 50\%$ of the explosion energy is
transferred to cosmic rays, reducing the X-ray emission by a factor of
20 from test-particle predictions. Efficiencies nearly this high are
found by Berezhko, Ksenofontov, \& Petukhov
(\cite{BereKP99a},\cite{BereKP99b}) and are typical of those shown
here in Table~1.

The four-component sum of the photon emission from the test-particle
result for SN1006 is shown as the heavy dotted curve in
Figure~\ref{fig:SN1006} where the emission volume has been adjusted to
fit the radio flux.  The broad-band fit is unsatisfactory and
does not improve substantially if parameters other than $\etainjP$ are
varied. In the test-particle result, the X-ray emission is
dominated by thermal \brems\ and the TeV to radio flux ratio is much
smaller than in the nonlinear result.

\section{SUMMARY AND CONCLUSIONS}
 \label{sec:conclusion}

We have developed a computationally fast and easy-to-use model of
nonlinear diffusive shock acceleration along with the accompanying
photon emission from the resultant electron and ion
spectra. Using this model, we showed how the emission depends on both
model and environmental parameters and identified ambient density and
magnetic field as being the most important in determining the
broad-band emission. We also showed that our simple model is in excellent
agreement with a more complete and complex model of
the supernova remnant SN1006.

A particularly important aspect of nonlinear acceleration is that
shock {\it heating} is linked to particle acceleration and thus to the
broad-band photon emission. As shown in Figures~\ref{fig:fpdjde},
\ref{fig:paraModel}a,
and \ref{fig:testpart}, shocks which accelerate particles efficiently
have lower postshock temperatures and higher postshock densities than
test-particle predictions. Nonlinear shocks also produce electron
spectra with
superthermal tails (Figures~\ref{fig:fpdjde}, \ref{fig:SNRcomp}, and
\ref{fig:testpart}).  These factors are
important for X-ray line emission  (\egc Dorfi \&
B\"ohringer \cite{DorfiB93}) and open up the possibility of using
radio, X-ray continuum, and $\gamma$-ray observations to constrain X-ray
line models and vice versa. We believe the model we present here is a
first step in this process.

To describe SNRs, we use a model of an expanding, spherical shock wave
(Truelove \& McKee \cite{Truelove99}) to obtain shock parameters as a
function of SNR parameters and time.  Using these parameters, we
calculate nonlinear particle soectra and produce ``snapshot''
continuum photon spectra from \syn, \brems, \IC, and \pion, spanning
the range from radio to TeV $\gamma$-rays.  More realistic models
would include inhomogeneous ejecta, emission from dense clumps, pre-SN
winds, a reverse shock, volume integrated emission from regions
undergoing adiabatic cooling, oblique shock geometry, and effects on
the SNR evolution from particle escape (or dilution).  Some of these
additions have already been performed by Berezhko, Ksenofontov,
\& Petukhov (\cite{BereKP99a}) and our results are in excellent
agreement with theirs for SN1006 (Figure~\ref{fig:SN1006} and
Section~\ref{sec:BereCompare}), some can be added easily to the simple
model, but others (such as oblique geometry) are much more difficult.
Despite this, we believe this nonlinear model is accurate enough to
approximate expected emission fluxes and clearly track important
trends as we describe in detail in Section~\ref{sec:keyparameters}.

Nonlinear shock acceleration unavoidably involves a large number of
model and environmental parameters and it is essential to know how
they relate to each other and which ones have the greatest impact on
the results.
Of the model parameters, the injection efficiency, $\etainjP$, has the
greatest influence on the solutions since it sets the overall
efficiency and determines whether the acceleration is nonlinear or can
be treated as test particle (Figures~\ref{fig:fpdjde} and
\ref{fig:paraModel}a). Unfortunately, $\etainjP$ is not well
constrained by theory and heliospheric observations remain limited; in
fact, there's virtually no constraining information on the efficiency
of high Mach number shocks like those expected in young SNRs. However,
the differences between test-particle and nonlinear predictions are so
large in both the broad-band continuum and the X-ray line emission,
that we believe it will be possible to set strong constraints on the
acceleration efficiency once nonlinear models of X-ray line emission
become available.
Other model parameters, such as the ratios of shocked electron to
proton temperature and of  electron to proton acceleration efficiency,
are less fundamental but produce characteristic differences in the
emission which can be constrained with sufficiently complete
models and observations (Figures~\ref{fig:paraModel}b and c).

The environmental parameters include the ambient interstellar proton
density $\ProDenUpS$ and density profile, the magnetic field $B_0$,
the ejecta mass and density profile, supernova explosion energy,
etc. Of these, the ambient density and magnetic field have, by far,
the largest effect on the broad-band emission. As we show in
Figures~\ref{fig:SNRcomp},
\ref{fig:paraInputSens}, and
\ref{fig:TeVtoRadio},
varying $\ProDenUpS$ and $B_0$ produce a complicated set of changes in the
emission which cannot be simply characterized. If density is held
constant, the intensity in the radio band scales roughly as
$B_0^{3/2}$, but at photon energies $E>$ MeV, the intensity can
decrease with increasing $B_0$ due to nonlinear effects
(Figure~\ref{fig:paraInputSens}b) and a weakening of the shock (\iec
decrease in the \alf\ Mach number).  Likewise, when $B_0$ is fixed, an
increase in density generally causes an increase in emission in the
MeV range, but the emission in the radio band is relatively
insensitive to density (Figure~\ref{fig:paraInputSens}a). At TeV
energies, the maximum photon energy increases as $\ProDenUpS$ is decreased
and, in general, the highest photon energies are obtained with the
largest $B_0$ and the lowest $\ProDenUpS$.
The TeV/radio flux ratio, however, is greatest for low ambient
magnetic fields (Figure~\ref{fig:TeVtoRadio}) and this is an important
parameter if radio SNRs are selected for hard $\gamma$-ray searches.  As we show
in Figure~\ref{fig:ic_pp},
the distinctive pion-decay bump will be most prominent for remnants
in high $\ProDenUpS$ and high $B_0$ environments;
given the flux levels predicted,
we anticipate positive detections of such spectral features (generally
below EGRET sensitivities for $n\lesssim 10$ cm$^{-3}$), by the GLAST
experiment in the not too distant future.  While TeV experiments
cannot probe such spectral features, they can detect the upper range
of pion decay emission in remnants in high $B_0$ surroundings; this
situation may already be realized in the recent announcement
(V\"olk et al. 2000) of a positive detection of Cas A by HEGRA.

The nature of the X-ray emission also depends importantly on density
and magnetic field. As density is increased at a given $B_0$, the keV
X-ray emission goes from being totally dominated by \syn\ from
relativistic electrons to quasi-thermal \brem\ emission.
Note
that the \brem\ dominates even in the radio band for $\ProDenUpS=10$ \pcc\
and $B_0=3$ \muG (Fig.~\ref{fig:paraInputSens}a).
Similarly, at a given density (Figure~\ref{fig:paraInputSens}b), the
X-ray emission goes from being thermal \brem\ to \syn\ as $B_0$
increases.  If \syn\ is dominant, X-ray lines will be weak or absent
so these differences are readily distinguishable observationally.

Finally, we suggest that the most important aspect of modeling photon
emission from astrophysical shocks depends on whether or not the
acceleration is efficient and nonlinear or inefficient and
test-particle in nature. These situations can be quite different and
if the acceleration is nonlinear, all parameters are interconnected
and observations in any energy band limit the freedom to vary
parameters in all other bands.  Since there are a large number of
parameters, understanding the nonlinear interactions requires a model
that can efficiently map parameter space.  We believe the model
presented here, while not a replacement for more complete models, can
do this expediently and accurately and be an aid in interpreting the
vast amount of information expected from current and future space and
ground-based telescopes.

\placefigure{fig:ic_pp}            

\vskip 1cm
\noindent
{\bf Acknowledgments:} We wish to thank John Raymond for providing
information concerning SN1006.  We also thank Rod Lessard for
providing Veritas flux sensitivity data and Anne Decourchelle for
helpful discussions.  This work was performed, in part, with support
from NASA's Space Physics Theory Program.

\vskip24pt


\back

\newpage

%
%
%

\clearpage

%
\hbox{}
\vskip-1.0truecm

\def\horule{\hrule height0.6pt}
\def\htrule{\hrule height0.8pt}
\newcommand\pc{\hbox{\%}}
\centerline{\rm TABLE~1\rm}
\centerline{\rm Parameters for Figure~\ref{fig:SNRcomp} and SN1006 \rm
}
\vskip-0.2truein
$$
\vbox{
\halign{\vphantom{$\big($}
\hfil$#$\hfil \ \ &
\hfil$#$\hfil &
\hfil$#$\hfil &
\hfil$#$\hfil &
\hfil$#$\hfil &
\hfil$#$\hfil &
\hfil$#$\hfil 
\cr
\noalign{\htrule\vskip1.5pt \htrule\vskip3.5pt}
\hbox{\ Input parameters \ \ \ } &
\hbox{\ \ \ \ Low \ \ \ } &
\hbox{\ \ \ \ High \ \ \ \ } &
\hbox{\ \ SN1006$_{\mathrm{NL}}$ \ \ } &
\hbox{\ \ BKP99$^{\bf 5}$ \ \ } &
\hbox{\ \ SN1006$_{\mathrm{TP}}$ \ \ } &
\cr
\noalign{\vskip 2pt\horule\vskip6pt}
\hbox{$\ProDenUpS$ [\pcc]$\,^{\bf 1}$} &
\hbox{$0.01$} &
\hbox{$10$} &
\hbox{$0.1$} &
\hbox{$0.1$} &
\hbox{$0.1$} &
\cr
\hbox{$B_0$ [\muG]} &
\hbox{$3$} &
\hbox{$30$} &
\hbox{$9$} &
\hbox{$9$} &
\hbox{$9$} &
\cr
\hbox{$B_2$ [\muG]$\,^{\bf 2}$} &
\hbox{$3$} &
\hbox{$30$} &
\hbox{$30$} &
\hbox{$30$} &
\hbox{$30$} &
\cr
\hbox{$\tSNR$ [yr]} &
\hbox{$1000$} &
\hbox{$1000$} &
\hbox{$990$} &
\hbox{$990$} &
\hbox{$990$} &
\cr
\hbox{$\EnSN$ [$10^{51}$ erg]} &
\hbox{$1$} &
\hbox{$1$} &
\hbox{$1$} &
\hbox{$1$} &
\hbox{$1$} &
\cr
\hbox{$\Mej$ [$M_{\sun}$]} &
\hbox{$1$} &
\hbox{$5$} &
\hbox{$1.4$} &
\hbox{$1.4$} &
\hbox{$1.4$} &
\cr
\hbox{$\TempPro$ [K]} &
\hbox{$10^4$} &
\hbox{$10^4$} &
\hbox{$10^4$} &
\hbox{$10^4$} &
\hbox{$10^4$} &
\cr
\hbox{$\TempRatio$} &
\hbox{$1$} &
\hbox{$1$} &
\hbox{$1$} &
\hbox{---} &
\hbox{$1$} &
\cr
\hbox{$\etainjP$$^{\bf 3}$} &
\hbox{$10^{-3}$} &
\hbox{$10^{-3}$} &
\hbox{$2.1\xx{-4}$} &
\hbox{$5\xx{-4}$} &
\hbox{$1\xx{-5}$} &
\cr
\hbox{$\epRatio$} &
\hbox{$0.03$} &
\hbox{$0.03$} &
\hbox{$2\xx{-3}$} &
\hbox{$2\xx{-3}$} &
\hbox{$2\xx{-3}$} &
\cr
\hbox{$\etamfp$} &
\hbox{$1$} &
\hbox{$1$} &
\hbox{$13$} &
\hbox{---} &
\hbox{$13$} &
\cr
\hbox{$\alpha$} &
\hbox{$1$} &
\hbox{$1$} &
\hbox{$0.6$} &
\hbox{$0.5$} &
\hbox{$0.6$} &
\cr
\hbox{$n$} &
\hbox{$0$} &
\hbox{$0$} &
\hbox{$7$} &
\hbox{$7$} &
\hbox{$7$} &
\cr
\noalign{\vskip 5.5pt\htrule\vskip6pt}
\hbox{Output values} &
\hbox{} &
\hbox{} &
\hbox{} &
\cr
\noalign{\vskip 5.5pt\htrule\vskip6pt}
\hbox{$\MSZ$} &
\hbox{$400$} &
\hbox{$85$} &
\hbox{$235$} &
\hbox{$170$} &
\hbox{$235$} &
\cr
\hbox{$\MAZ$} &
\hbox{$120$} &
\hbox{$77$} &
\hbox{$72$} &
\hbox{$67$} &
\hbox{$72$} &
\cr
\hbox{$\Vsk$ [\kmps ]} &
\hbox{$6400$} &
\hbox{$1350$} &
\hbox{$3750$} &
\hbox{$3520$} &
\hbox{$3750$} &
\cr
\hbox{$\Rsk$ [pc]} &
\hbox{$11$} &
\hbox{$3.0$} &
\hbox{$7.0$} &
\hbox{$6.9$} &
\hbox{$7.0$} &
\cr
\hbox{$\Rtot$ } &
\hbox{$9.7$} &
\hbox{$8.5$} &
\hbox{$7.2$} &
\hbox{$6.7$} &
\hbox{$4.1$} &
\cr
\hbox{$\Rsub$ } &
\hbox{$3.0$} &
\hbox{$2.9$} &
\hbox{$3.6$} &
\hbox{$3.6$} &
\hbox{$4.0$} &
\cr
\hbox{$\EmaxPro$ [TeV] } &
\hbox{$30$} &
\hbox{$16$} &
\hbox{$9$} &
\hbox{$10$} &
\hbox{$11$} &
\cr
\hbox{$\EmaxElec$ [TeV] } &
\hbox{$30$} &
\hbox{$12$} &
\hbox{$9$} &
\hbox{$10$} &
\hbox{$10$} &
\cr
\hbox{$\etainjE$ } &
\hbox{$5\xx{-3}$} &
\hbox{$0.11$} &
\hbox{$2.9\xx{-5}$} &
\hbox{---} &
\hbox{$8\xx{-7}$} &
\cr
\hbox{$\DStemp$ [K]} &
\hbox{$6.4\xx{7}$} &
\hbox{$3.6\xx{6}$} &
\hbox{$4.7\xx{7}$} &
\hbox{---} &
\hbox{$1.7\xx{8}$} &
\cr
\hbox{$\DStp$ [K]$^{\bf 4}$} &
\hbox{$5.0\xx{8}$} &
\hbox{$2.2\xx{7}$} &
\hbox{$1.7\xx{8}$} &
\hbox{---} &
\hbox{$1.7\xx{8}$} &
\cr
\hbox{$\EffRel$} &
\hbox{$0.80$} &
\hbox{$0.75$} &
\hbox{$0.63$} &
\hbox{$0.40$} &
\hbox{$0.02$} &
\cr
\noalign{\vskip 5.5pt\htrule\vskip6pt}
\hbox{Flux parameters} &
\hbox{} &
\hbox{} &
\hbox{} &
\cr
\noalign{\vskip 5.5pt\htrule\vskip6pt}
\hbox{$\dSNR$ [kpc]} &
\hbox{$1$} &
\hbox{$1$} &
\hbox{$1.7$} &
\hbox{$1.7$} &
\hbox{$1.7$} &
\cr
\hbox{$\EmisVol$ [pc$^3$]} &
\hbox{$1$} &
\hbox{$1$} &
\hbox{$195$$\,^{\bf 6}$} &
\hbox{---} &
\hbox{$5\xx{3}$$\,^{\bf 7}$} &
\cr
\noalign{\vskip 2.5pt\htrule} } }
$$

$\,^{\bf 1}$ This is the unshocked proton density. In all cases,
including our match to
the \BKP\ example, a helium number density of $0.1\, \ProDenUpS$ is
assumed. 

$\,^{\bf 2}$ In order to compare our model to that of \BKP\
(\cite{BereKP99a}), we include a compressed downstream magnetic field
for purposes of calculating the \syn\ emission from SN1006. In all 
other results we take $B_2 = B_0$.

$\,^{\bf 3}$ See the discussion in Section~\ref{sec:BereCompare}
concerning this parameter and the \BKP\ results.

$\,^{\bf 4}$ This is the temperature of the shocked gas in the
test-particle approximation. 

$\,^{\bf 5}$ The parameters given here are for the model shown as a
solid line in Figure~1d of Berezhko, Ksenofontov,
\& Petukhov (\cite{BereKP99a}).

$\,^{\bf 6}$ This value is that given in
Eqn.~(\ref{eq:Emission}).

$\,^{\bf 7}$ This value is $\sim 15$ times $\EmisVol$ given in
Eqn.~(\ref{eq:Emission}).

%

\newpage

\textwidth=7.5truein \hoffset-0.25truein

%
\hbox{}
\vskip-1.0truecm

\centerline{\rm TABLE~2\rm}
\centerline{\rm Parameters for Figure~\ref{fig:paraInputSens} \rm
}
\vskip-0.2truein
$$
\vbox{
\halign{\vphantom{$\big($}
\hfil$#$\hfil \ \ &
\hfil$#$\hfil &
\hfil$#$\hfil &
\hfil$#$\hfil &
\hfil$#$\hfil &
\hfil$#$\hfil &
\hfil$#$\hfil &
\hfil$#$\hfil &
\hfil$#$\hfil &
\hfil$#$\hfil &
\hfil$#$\hfil &
\hfil$#$\hfil &
\hfil$#$\hfil \cr
\noalign{\htrule\vskip1.5pt \htrule\vskip3.5pt}
\hbox{} &
\hbox{} &
\hbox{(a)} &
\hbox{} &
\hbox{} &
\hbox{} &
\hbox{(b)} &
\cr
\hbox{\ Input \ \ } &
\hbox{\ \ solid \ \ } &
\hbox{\ \ dashed \ \ } &
\hbox{dot-dashed \ \ } &
\hbox{dotted \ \ } &
\hbox{\ \ \ \ \ solid \ \ } &
\hbox{\ \ dashed \ \ } &
\hbox{dot-dashed \ \ } &
\hbox{dotted \ \ } &
\cr
\noalign{\vskip 2pt\horule\vskip6pt}
\hbox{$\ProDenUpS$ [\pcc]} &
\hbox{$0.01$} &
\hbox{$0.1$} &
\hbox{$1$} &
\hbox{$10$} &
\hbox{\ \ $1$} &
\hbox{$1$} &
\hbox{$1$} &
\hbox{$1$} &
\cr
\hbox{$B_0$ [\muG]} &
\hbox{$3$} &
\hbox{$3$} &
\hbox{$3$} &
\hbox{$3$} &
\hbox{\ \ $3$} &
\hbox{$10$} &
\hbox{$30$} &
\hbox{$100$} &
\cr
\hbox{$\tSNR$ [yr]} &
\hbox{$1000$} &
\hbox{$1000$} &
\hbox{$1000$} &
\hbox{$1000$} &
\hbox{\ \ $1000$} &
\hbox{$1000$} &
\hbox{$1000$} &
\hbox{$1000$} &
\cr
\hbox{$\EnSN$ [$10^{51}$ erg]} &
\hbox{$1$} &
\hbox{$1$} &
\hbox{$1$} &
\hbox{$1$} &
\hbox{\ \ $1$} &
\hbox{$1$} &
\hbox{$1$} &
\hbox{$1$} &
\cr
\hbox{$\Mej$ [$M_{\sun}$]} &
\hbox{$1$} &
\hbox{$1$} &
\hbox{$1$} &
\hbox{$1$} &
\hbox{\ \ $1$} &
\hbox{$1$} &
\hbox{$1$} &
\hbox{$1$} &
\cr
\hbox{$\TempPro$ [K]} &
\hbox{$10^4$} &
\hbox{$10^4$} &
\hbox{$10^4$} &
\hbox{$10^4$} &
\hbox{\ \ $10^4$} &
\hbox{$10^4$} &
\hbox{$10^4$} &
\hbox{$10^4$} &
\cr
\hbox{$\TempRatio$} &
\hbox{$1$} &
\hbox{$1$} &
\hbox{$1$} &
\hbox{$1$} &
\hbox{\ \ $1$} &
\hbox{$1$} &
\hbox{$1$} &
\hbox{$1$} &
\cr
\hbox{$\etainjP$} &
\hbox{$10^{-3}$} &
\hbox{$10^{-3}$} &
\hbox{$10^{-3}$} &
\hbox{$10^{-3}$} &
\hbox{\ \ $10^{-3}$} &
\hbox{$10^{-3}$} &
\hbox{$10^{-3}$} &
\hbox{$10^{-3}$} &
\cr
\hbox{$\epRatio$} &
\hbox{$0.03$} &
\hbox{$0.03$} &
\hbox{$0.03$} &
\hbox{$0.03$} &
\hbox{\ \ $0.03$} &
\hbox{$0.03$} &
\hbox{$0.03$} &
\hbox{$0.03$} &
\cr
\noalign{\vskip 5.5pt\htrule\vskip6pt}
\hbox{Output} &
\hbox{} &
\hbox{} &
\hbox{} &
\cr
\noalign{\vskip 5.5pt\htrule\vskip6pt}
\hbox{$\MSZ$} &
\hbox{$400$} &
\hbox{$220$} &
\hbox{$130$} &
\hbox{$79$} &
\hbox{\ \ $130$} &
\hbox{$130$} &
\hbox{$130$} &
\hbox{$130$} &
\cr
\hbox{$\MAZ$} &
\hbox{$110$} &
\hbox{$200$} &
\hbox{$370$} &
\hbox{$720$} &
\hbox{\ \ $370$} &
\hbox{$110$} &
\hbox{$37$} &
\hbox{$11$} &
\cr
\hbox{$\Vsk$ [\kmps ]} &
\hbox{$6340$} &
\hbox{$3440$} &
\hbox{$2050$} &
\hbox{$1260$} &
\hbox{\ \ $2050$} &
\hbox{$2050$} &
\hbox{$2050$} &
\hbox{$2050$} &
\cr
\hbox{$\Rsk$ [pc]} &
\hbox{$10.5$} &
\hbox{$7.4$} &
\hbox{$4.8$} &
\hbox{$3.1$} &
\hbox{\ \ 4.8} &
\hbox{4.8} &
\hbox{4.8} &
\hbox{4.8} &
\cr
\hbox{$\Rtot$ } &
\hbox{$9.6$} &
\hbox{$12$} &
\hbox{$14$} &
\hbox{$17$} &
\hbox{\ \ $14$} &
\hbox{$9.6$} &
\hbox{$6.7$} &
\hbox{$4.8$} &
\cr
\hbox{$\Rsub$ } &
\hbox{$3.0$} &
\hbox{$3.0$} &
\hbox{$3.0$} &
\hbox{$3.0$} &
\hbox{\ \ $3.0$} &
\hbox{$2.9$} &
\hbox{$3.0$} &
\hbox{$3.2$} &
\cr
\hbox{$\EmaxPro$ [TeV] } &
\hbox{$32$} &
\hbox{$8$} &
\hbox{$2.4$} &
\hbox{$0.8$} &
\hbox{\ \ $2.4$} &
\hbox{$11$} &
\hbox{$44$} &
\hbox{$180$} &
\cr
\hbox{$\EmaxElec$ [TeV] } &
\hbox{$32$} &
\hbox{$8$} &
\hbox{$2.4$} &
\hbox{$0.8$} &
\hbox{\ \ $2.4$} &
\hbox{$11$} &
\hbox{$20$} &
\hbox{$12$} &
\cr
\hbox{$\etainjE$ } &
\hbox{$5\xx{-3}$} &
\hbox{$0.035$} &
\hbox{$6\xx{-3}$} &
\hbox{$7\xx{-3}$} &
\hbox{\ \ $6\xx{-3}$} &
\hbox{$9\xx{-3}$} &
\hbox{$0.011$} &
\hbox{$0.013$} &
\cr
\hbox{$\DStemp$ [K]} &
\hbox{$6.3\xx{7}$} &
\hbox{$1.3\xx{7}$} &
\hbox{$3.0\xx{6}$} &
\hbox{$7.8\xx{5}$} &
\hbox{\ \ $3.0\xx{6}$} &
\hbox{$6.6\xx{6}$} &
\hbox{$1.3\xx{7}$} &
\hbox{$2.8\xx{7}$} &
\cr
\hbox{$\DStp$ [K]} &
\hbox{$4.9\xx{8}$} &
\hbox{$1.5\xx{8}$} &
\hbox{$5.1\xx{7}$} &
\hbox{$1.9\xx{7}$} &
\hbox{\ \ $5.1\xx{7}$} &
\hbox{$5.1\xx{7}$} &
\hbox{$5.1\xx{7}$} &
\hbox{$5.1\xx{7}$} &
\cr
\hbox{$\EffRel$} &
\hbox{$0.81$} &
\hbox{$0.85$} &
\hbox{$0.88$} &
\hbox{$0.90$} &
\hbox{\ \ $0.88$} &
\hbox{$0.79$} &
\hbox{$0.63$} &
\hbox{$0.37$} &
\cr
\noalign{\vskip 5.5pt\htrule\vskip6pt}
\hbox{Flux} &
\hbox{} &
\hbox{} &
\hbox{} &
\cr
\noalign{\vskip 5.5pt\htrule\vskip6pt}
\hbox{$\dSNR$ [kpc]} &
\hbox{$1$} &
\hbox{$1$} &
\hbox{$1$} &
\hbox{$1$} &
\hbox{\ \ $1$} &
\hbox{$1$} &
\hbox{$1$} &
\hbox{$1$} &
\cr
\hbox{$\EmisVol$ [pc$^3$]} &
\hbox{$510$} &
\hbox{$140$} &
\hbox{$33$} &
\hbox{$7.2$} &
\hbox{\ \ $33$} &
\hbox{$50$} &
\hbox{$70$} &
\hbox{$100$} &
\cr
\noalign{\vskip 2.5pt\htrule} } }
$$
%

\newpage

\textwidth=6.5truein \hoffset-0.0truein

\figureoutsmall{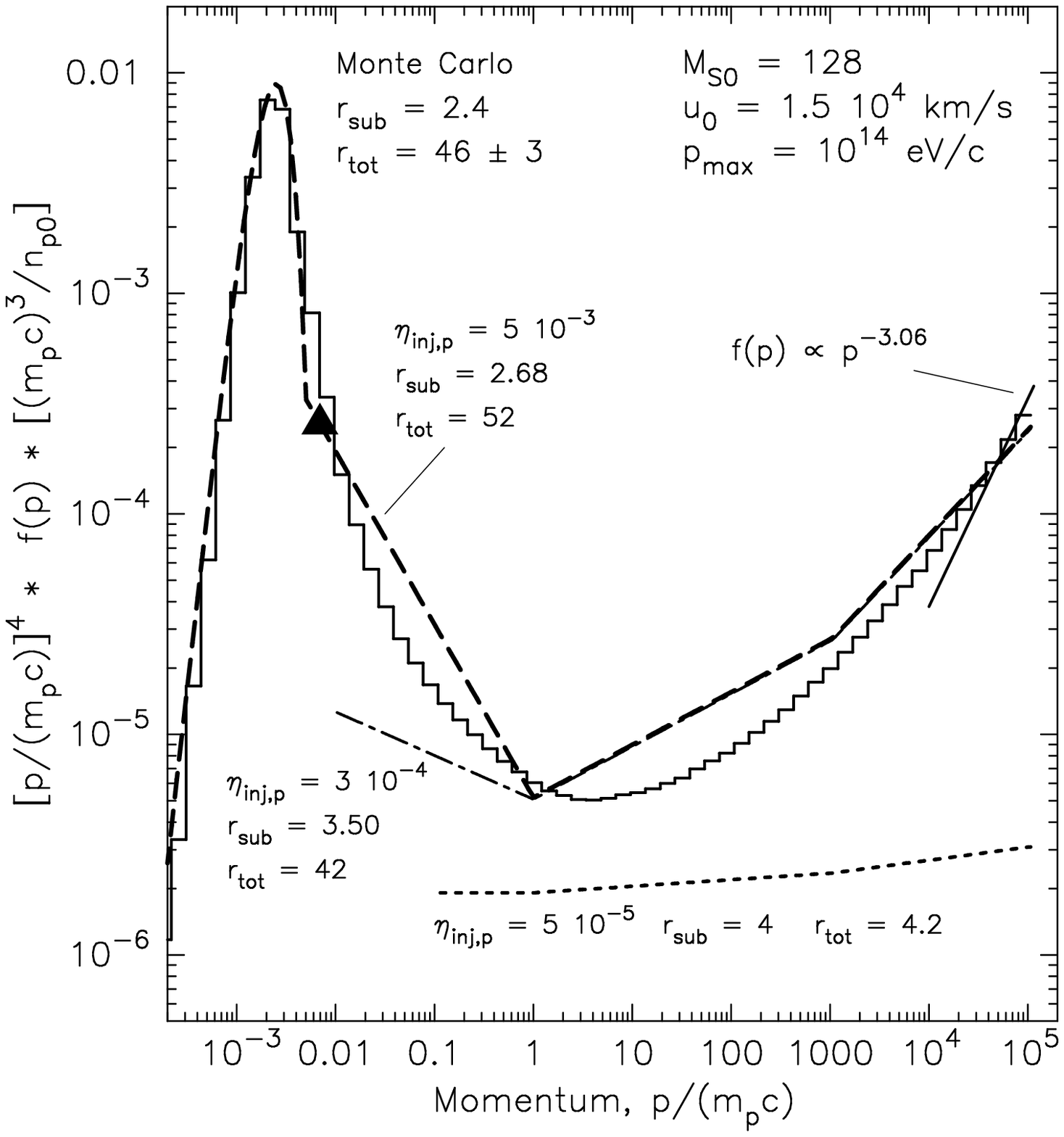}{
Downstream phase space distribution functions, $f$, versus momentum,
$p$. We have multiplied $f(p)$ by $[p/(\mpc)]^4$ to flatten the
spectra, and by $[(\mpc)^3/\ProDenUpS]$ to make them dimensionless.  The
solid histogram is a Monte Carlo model result, while the dashed,
dot-dashed, and dotted curves are from the simple model. The three
simple model results are obtained with identical input parameters
except for $\etainjP$ which is varied as shown. The $\etainjP=5\xx{-3}$
case is chosen to match the \MC\ result and a good correspondence
between the two models is obtained above the injection momentum,
$\pinj \simeq 7\xx{-3} \, \mpc $. The injection momentum, $\pinj$,
varies with input parameters: for $\etainjP=3\xx{-4}$, $\pinj \simeq
1.0\xx{-2} \, \mpc $, and for $\etainjP=5\xx{-5}$, $\pinj \simeq 0.11 \,
\mpc $.  As explained in Berezhko \& Ellison (1999), the model spectra
are considerably steeper than the test-particle prediction
for
$r_{\rm tot}=46$, $f \propto
p^{-3.06}$ (solid line), at the highest momenta.
\label{fig:EBfive} }

\figureoutvsmall{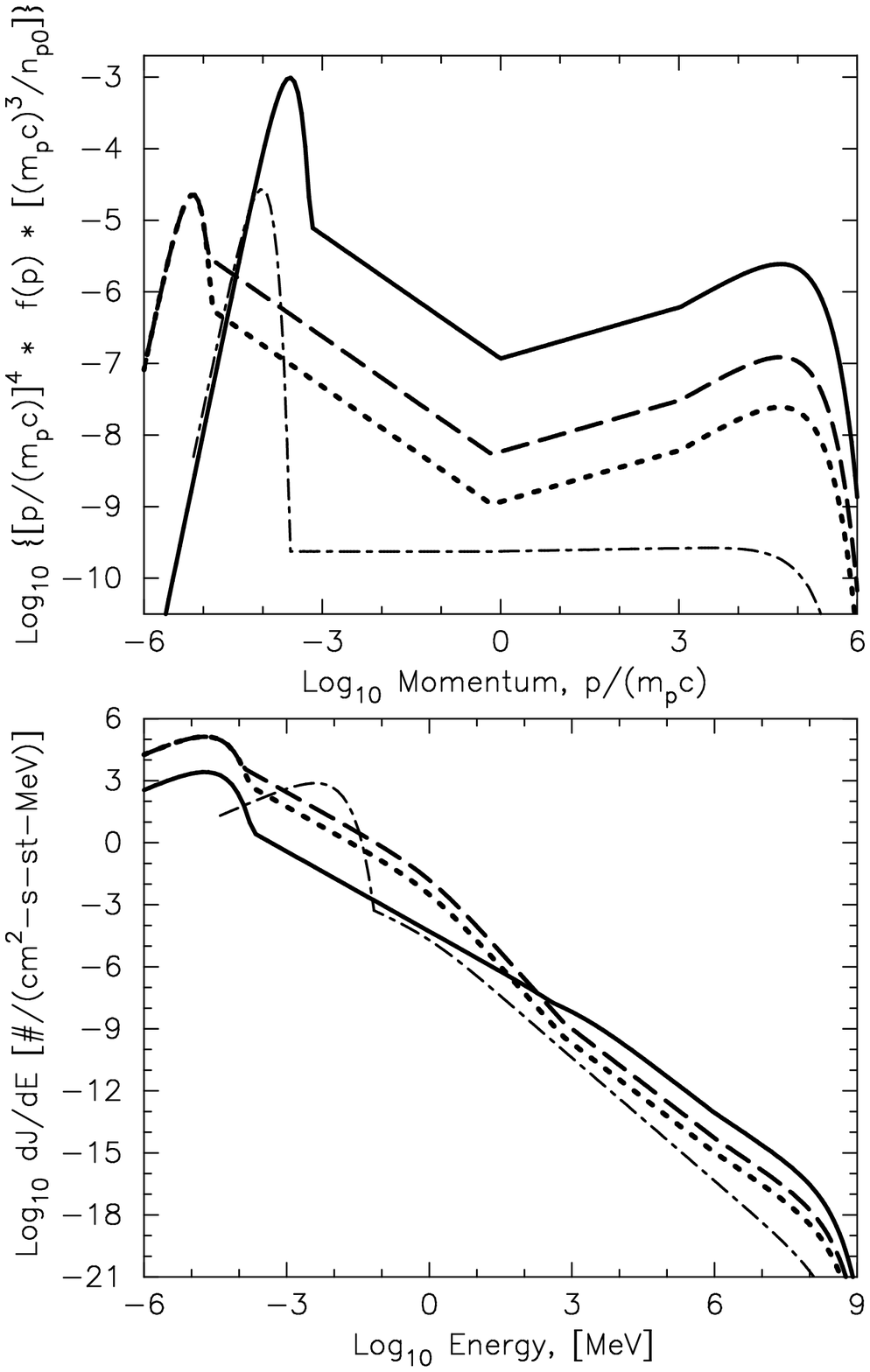}{
Top panel: Phase space distribution functions, $f(p)$, flattened
and made dimensionless as in Fig.~\ref{fig:EBfive}. The solid curve
shows protons, the dashed curve shows electrons with $\epRatio=0.05$,
and the dotted curve shows electrons with $\epRatio=0.01$. No
\syn\ or \IC\ losses are included so the electron and proton spectra
cutoff at the same momentum. Bottom panel: Differential energy flux
distributions for the same particles. These spectral are normalized
with a far upstream number density, $n^{\prime}_0$,
such that $n^{\prime}_0 \, u_0 = 1$ cm$^{-2}$ s$^{-1}$. The dot-dashed
curves are test-particle electron results with $\epRatio=0.01$ included for
comparison. All input parameters are the same as for the dotted curves
except
$\etainjP=5\xx{-6}$ which gives $\Rtot=4.1$.
\label{fig:fpdjde}  }

\figureoutvsmall{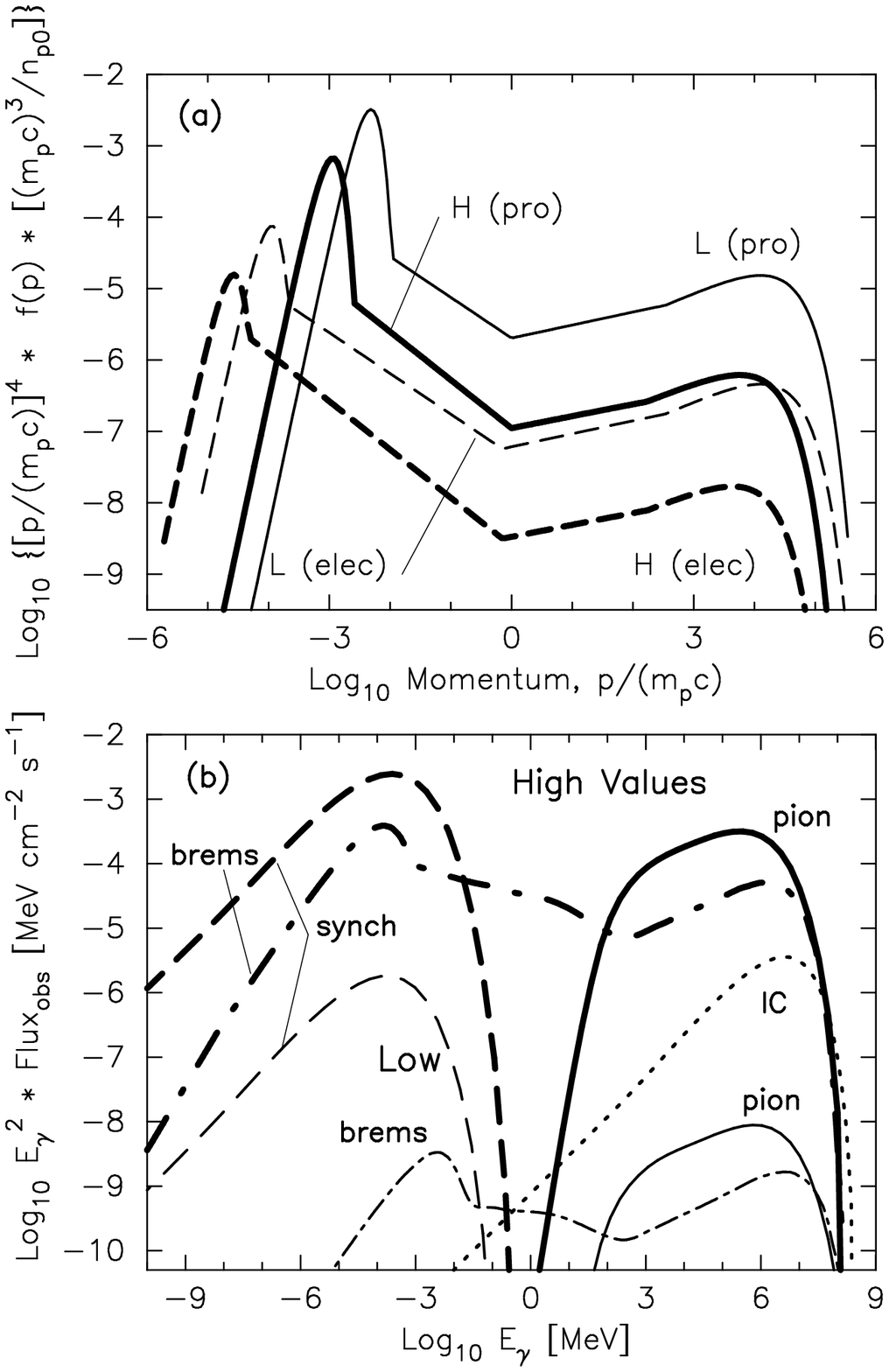}{
Top panel: Phase space distributions for a parameter set with low
values of $\ProDenUpS$, $\Mej$, and $B_0$
(light-weight curves) and high values
(heavy-weight curves). The protons are shown as solid curves and
the electrons are shown as dashed curves. Bottom panel: Photon fluxes
at Earth from the distributions in the top panel assuming an emission
volume of $\EmisVol = 1$ pc$^{3}$ at a distance of 1 kpc. In each
case, the \syn\ emission is shown with a dashed curve, the \brem\ with
a dot-dashed curve, the \IC\ with a dotted curve, and the \pion\ with
a solid curve. The \IC\ emission for the high value example is
insignificant and is not shown. The full set of parameters for these
models is given in Table~1 under columns marked `High' and `Low.'
\label{fig:SNRcomp}  }

\figureoutvsmall{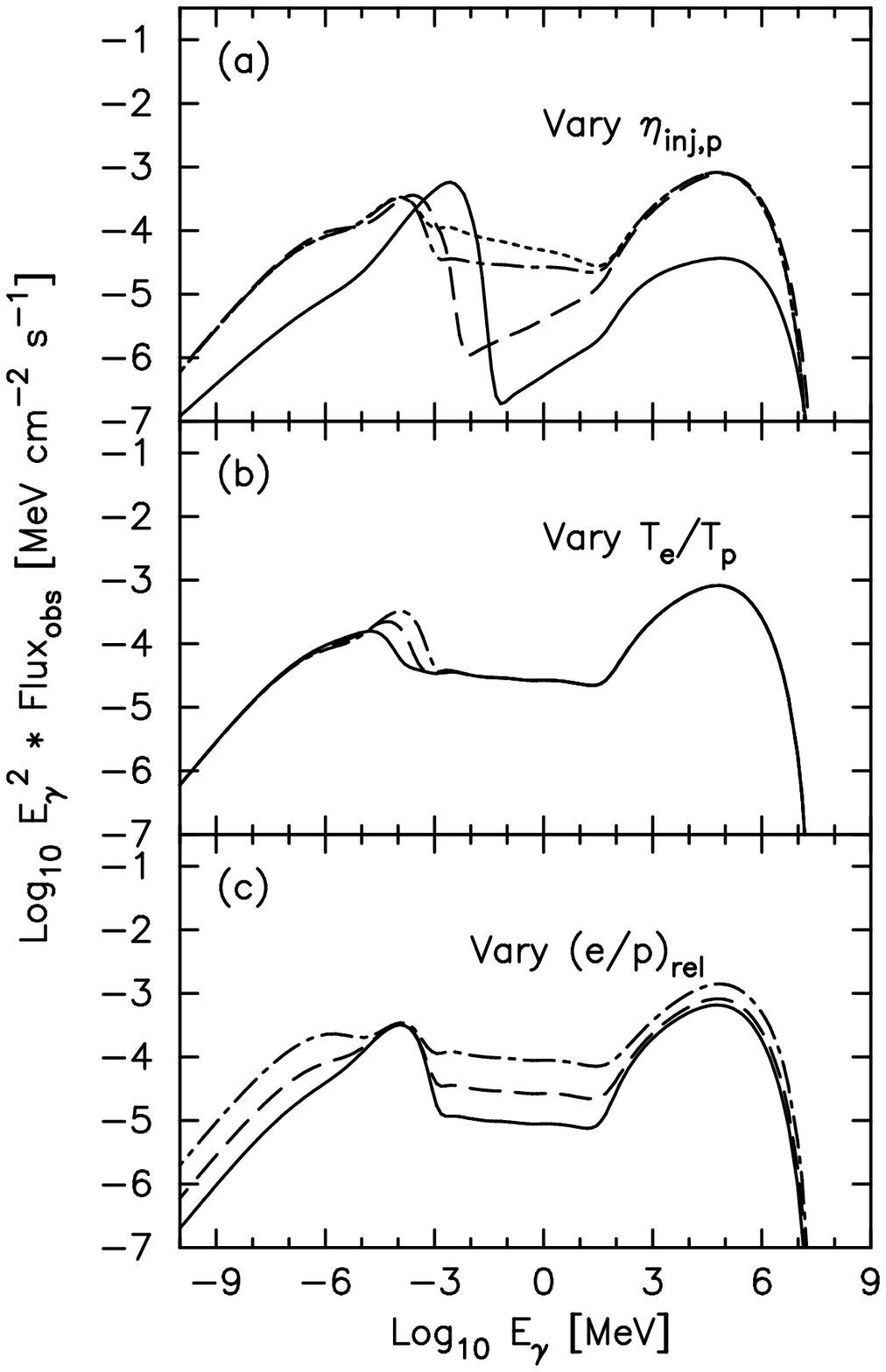}{
Total photon emission for various model parameters. In (a),
$\etainjP=1\xx{-5}$ (solid),
$1\xx{-4}$ (dashed),
$1\xx{-3}$ (dot-dashed), and
$2\xx{-3}$ (dotted) all with
$\TempRatio = 1$ and $\epRatio = 0.03$.
In (b):
$\TempRatio=0.1$ (solid),
$0.3$ (dashed), and
$1$ (dot-dashed) all with
$\etainjP=1\xx{-3}$ and
$\epRatio=0.03$.
In (c):
$\epRatio=0.01$ (solid),
$0.03$ (dashed), and
$0.1$ (dot-dashed)
all with
$\etainjP=1\xx{-3}$ and $\TempRatio=1$.
In all cases,
$\ProDenUpS=1$ \pcc,
$B_0=3$ \muG,
$\EnSN=1\xx{51}$ erg,
$\Mej=1 M_{\sun}$,
and
$\tSNR=10^3$ yr.
Note that the vertical axis is proportional to the standard $\nu \,
F_\nu$ representation.
\label{fig:paraModel} }

\figureout{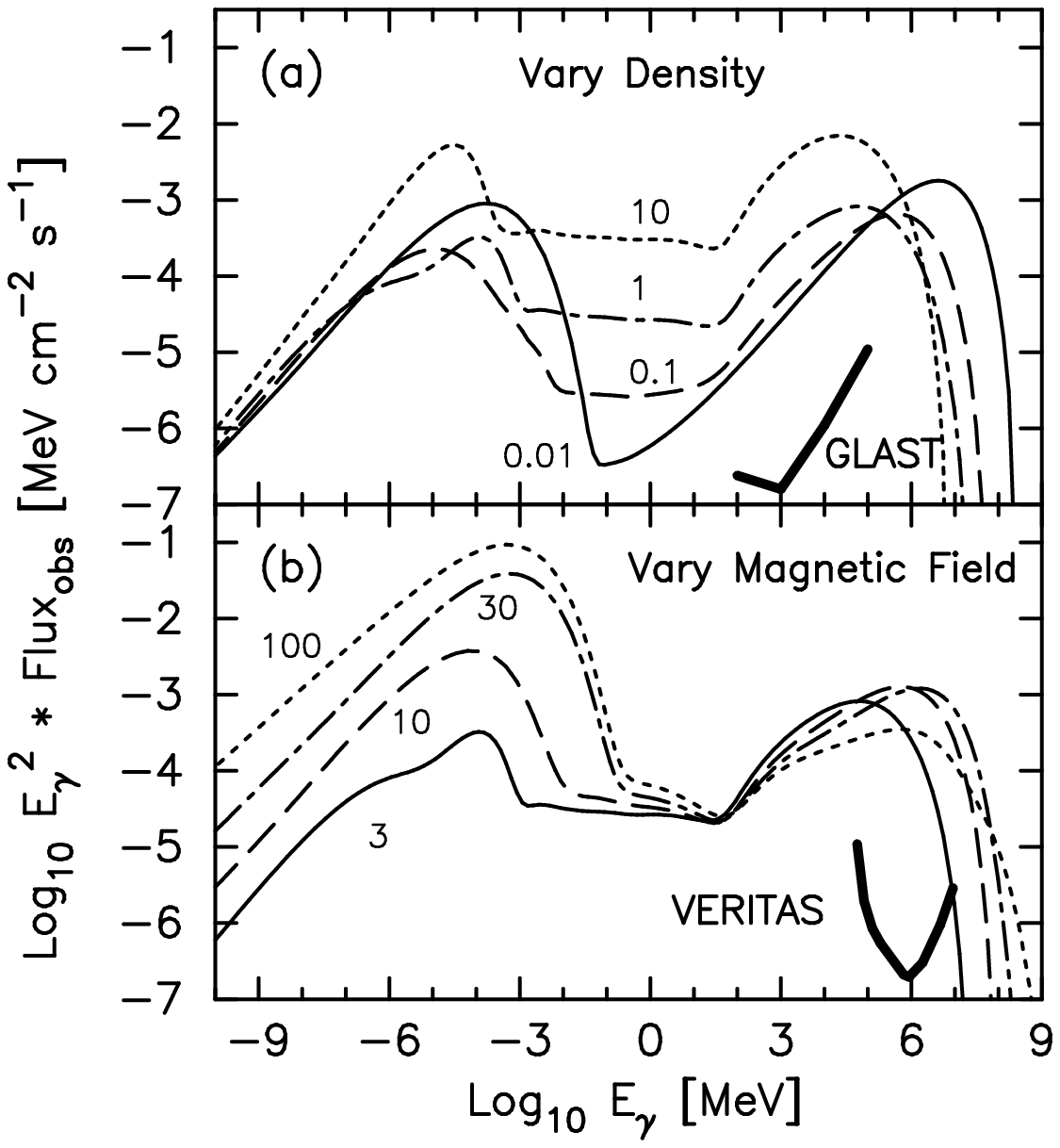}{
Total photon emission for various input parameters. In (a), the
ISM field is fixed at $B_0=3$ \muG\ and the ambient number density is
varied such that:
$\ProDenUpS=0.01$ \pcc\ (solid),
$0.1$ \pcc\ (dashed),
$1$ \pcc\ (dot-dashed), and
$10$ \pcc\ (dotted).
In (b), $B_0$ is varied:
$B_0=3$ \muG\ (solid),
$10$ \muG\ (dashed),
$30$ \muG\ (dot-dashed), and
$100$ \muG\ (dotted), with the density pinned to $\ProDenUpS=1$ \pcc.
The input and output parameters for these examples are given in
Table~2, but in all cases,
$\etainjP=10^{-3}$,
$\TempRatio=1$,
$\epRatio=0.03$,
$\EnSN=10^{51}$ erg,
$\Mej=1 M_{\sun}$,
$\tSNR=1000$ yr, 
$\alpha=1$,
$\etamfp=1$,
and the flux at Earth is determined assuming
$\dSNR=1$ kpc and an emission volume given by Eqn.~(\ref{eq:Emission}).
Also depicted are the canonical integral flux sensitivities for Veritas
and GLAST (\egc Weekes \etal\ 1999).
\label{fig:paraInputSens}  }

\figureout{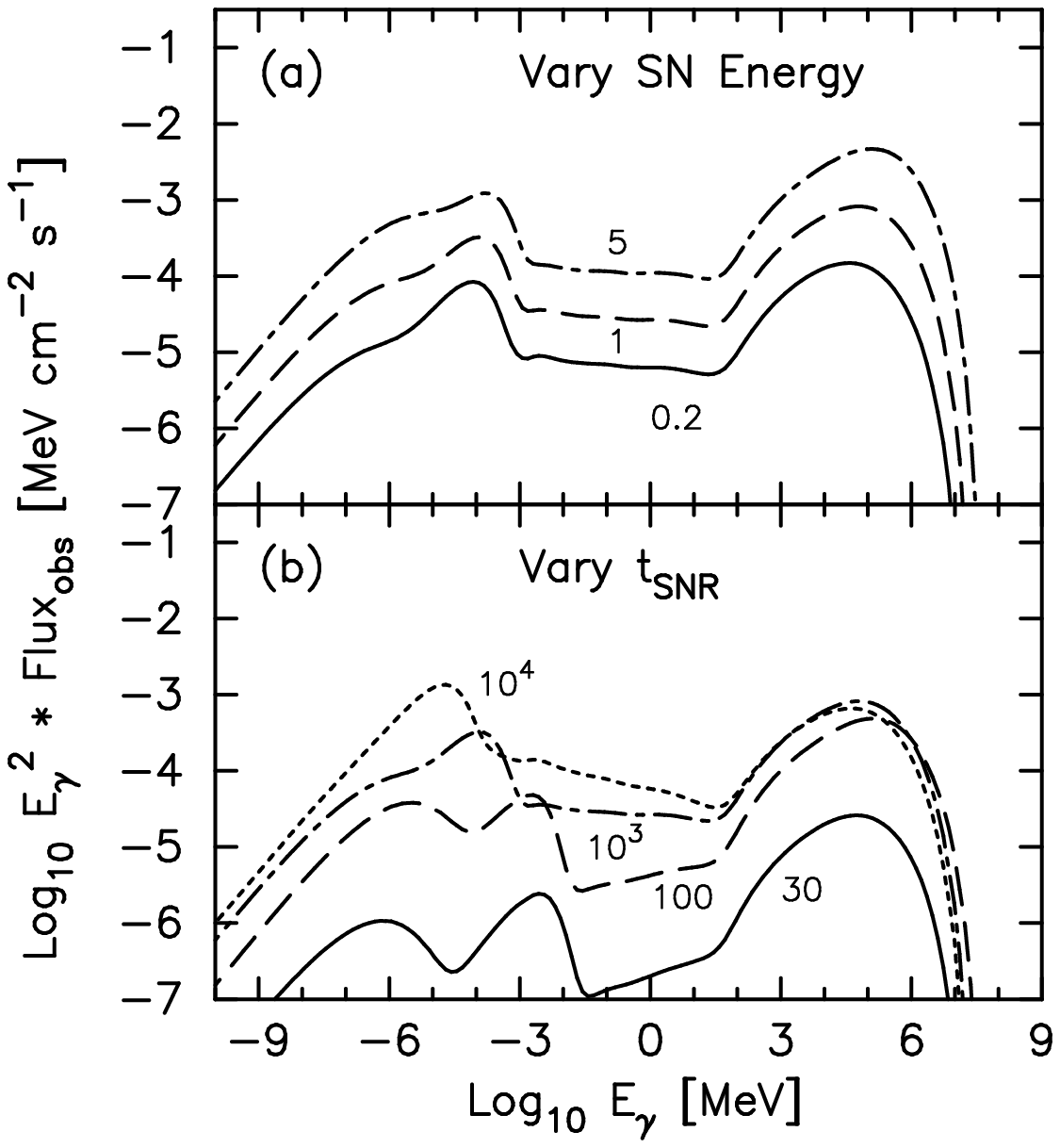}{
Total photon emission for various input parameters. In
(a),
$\EnSN = 0.2\xx{51}$ erg (solid),
$1\xx{51}$ erg (dashed), and
$5\xx{51}$ erg (dot-dashed).
In (b),
$\tSNR = 30$ yr (solid),
$100$ yr (dashed),
$10^3$ yr (dot-dashed), and
$10^4$ yr (dotted).
In all cases,
$\etainjP=10^{-3}$,
$\ProDenUpS=1$ \pcc,
$B=3$\muG,
$\TempRatio=1$,
$\epRatio=0.03$,
$\Mej=1 M_{\sun}$, 
$\alpha=1$,
$\etamfp=1$,
and the flux at Earth is determined assuming
$\dSNR=1$ kpc and an emission volume given by Eqn.~(\ref{eq:Emission}).
\label{fig:paraInputBot}  }

\figureout{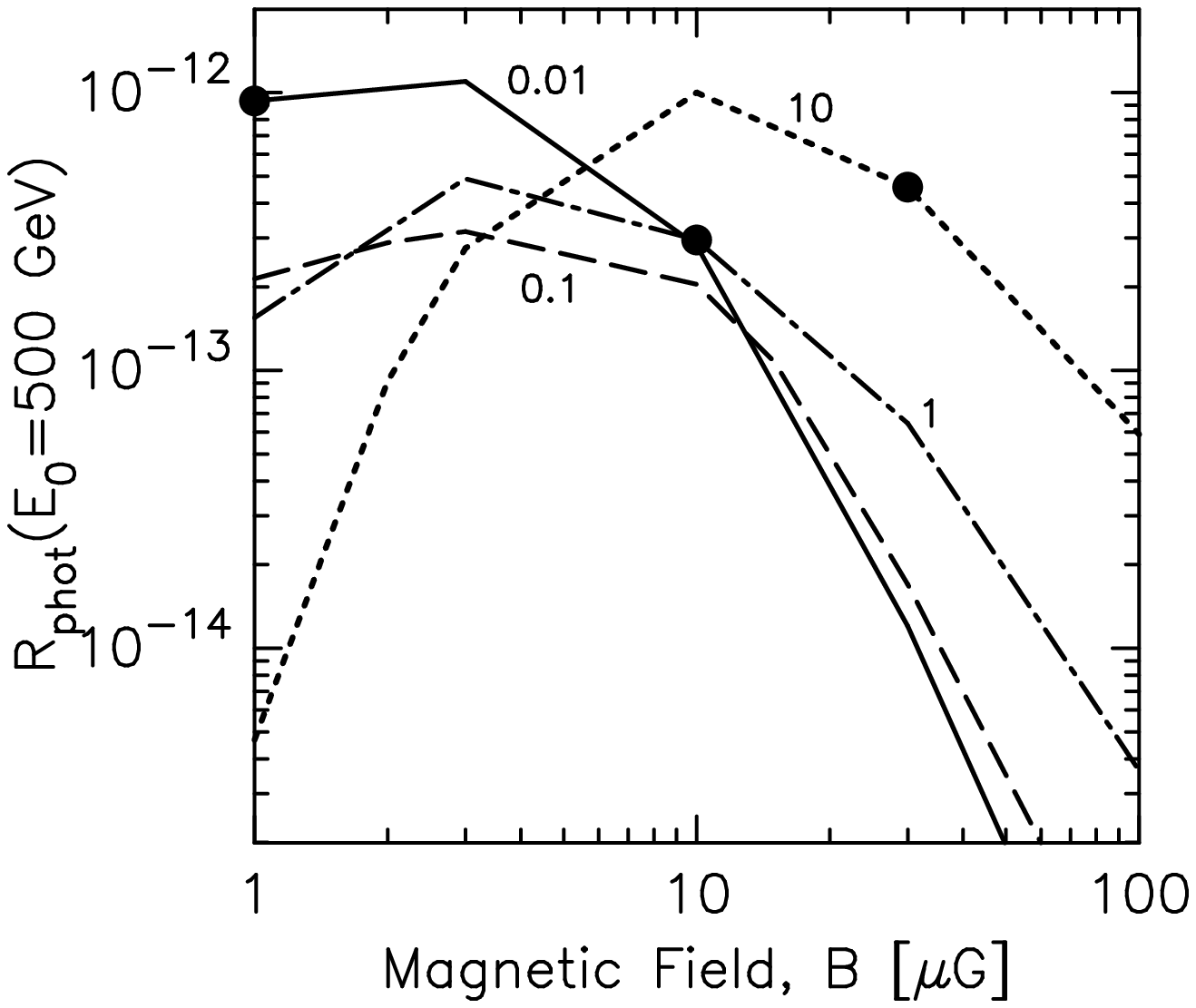}{
The ratio of gamma-ray to radio fluxes, $\Rphot$, as defined in the
text as a function of magnetic field and unshocked density as marked
in \pcc.
All plots have
$\EnSN=10^{51}$ erg,
$\Mej=1 M_{\sun}$,
$\epRatio = 0.03$, 
$\etainjP=10^{-3}$, 
$\tSNR=1000$ yr, 
$\alpha=1$,
$\etamfp=1$,
and
$\TempRatio = 1$.
The dots indicate parameters used in Figure~\ref{fig:ic_pp}.
\label{fig:TeVtoRadio}  }

\figureoutvsmall{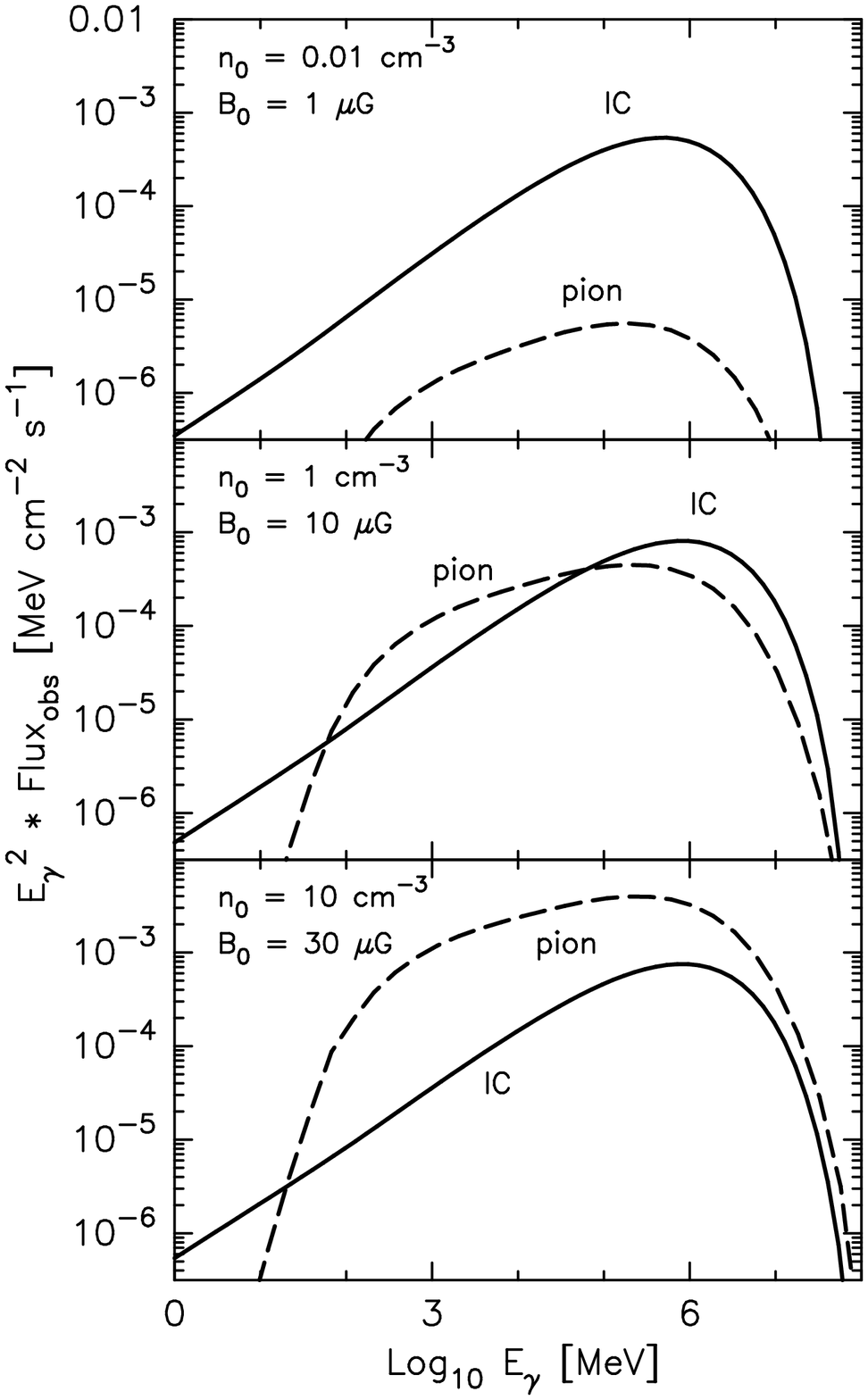}{
Inverse-Compton and \pion\  components for the densities and magnetic fields
marked with  dots in Fig.~\ref{fig:TeVtoRadio}. 
\label{fig:ic_pp}  }

\figureoutsmall{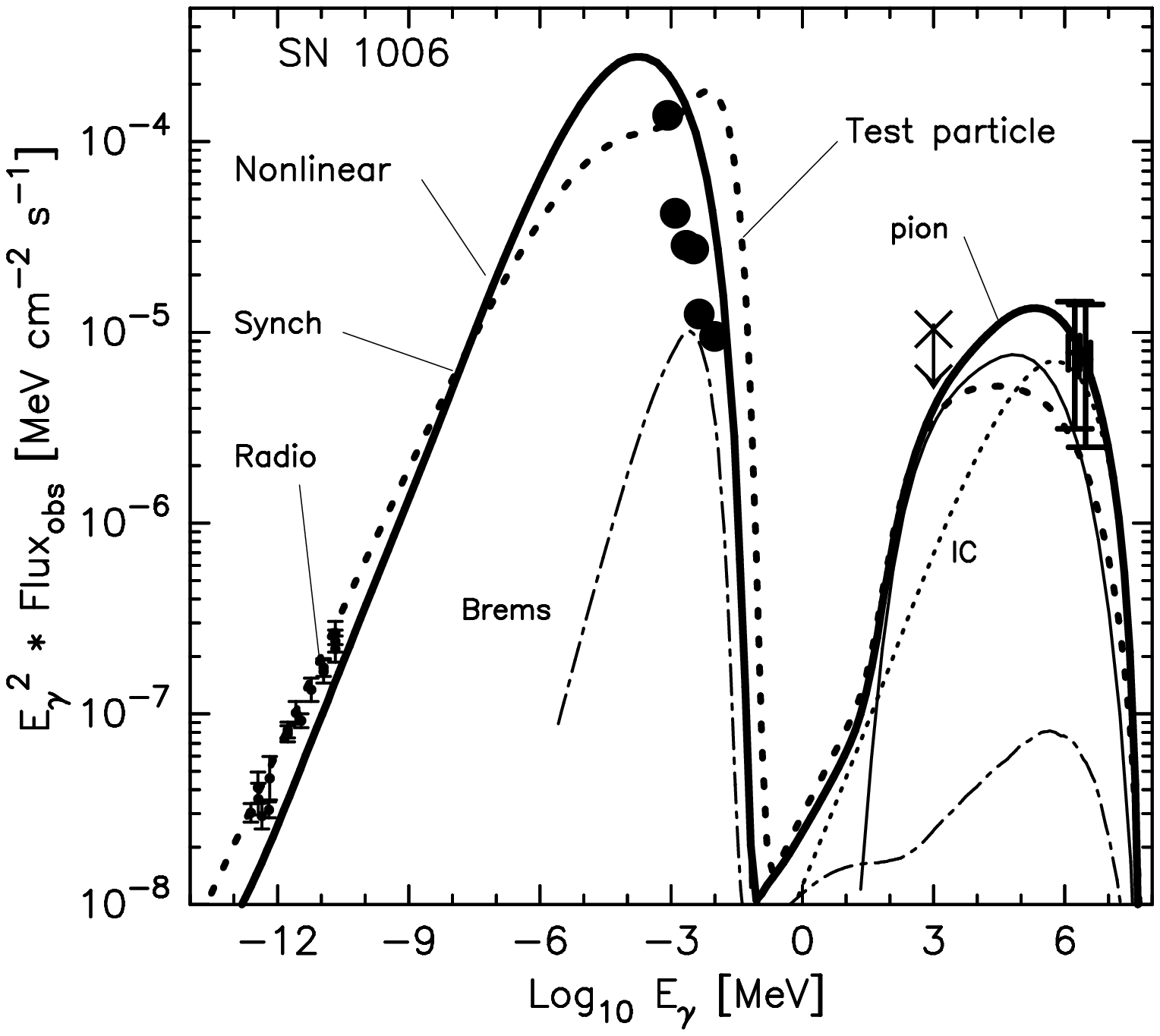}{
Photon spectra for SN1006. All spectra are calculated at the shock and
show a `snapshot' of the remnant at 990 yr after the explosion.  The
heavy-weight solid and dotted curves are sums of the four photon
components. The component spectra for the heavy-weight solid curve are
shown and labeled.  The heavy solid curve is the result for a
nonlinear shock, while the heavy dotted curve is a test-particle
result.  The radio data are from Reynolds \& Ellison (\cite{RE92}) and
the X-ray data (solid points) are adapted from Reynolds
(\cite{Reynolds96}), the EGRET upper limit (cross) is from
Mastichiadis \& de Jager (1996), and the CANGAROO TeV points (squares)
are from Tanimori \etal (1998). Note that the integral CANGAROO points
are plotted, with no adjustment, on our differential representation.
\label{fig:SN1006}  }

\figureoutsmall{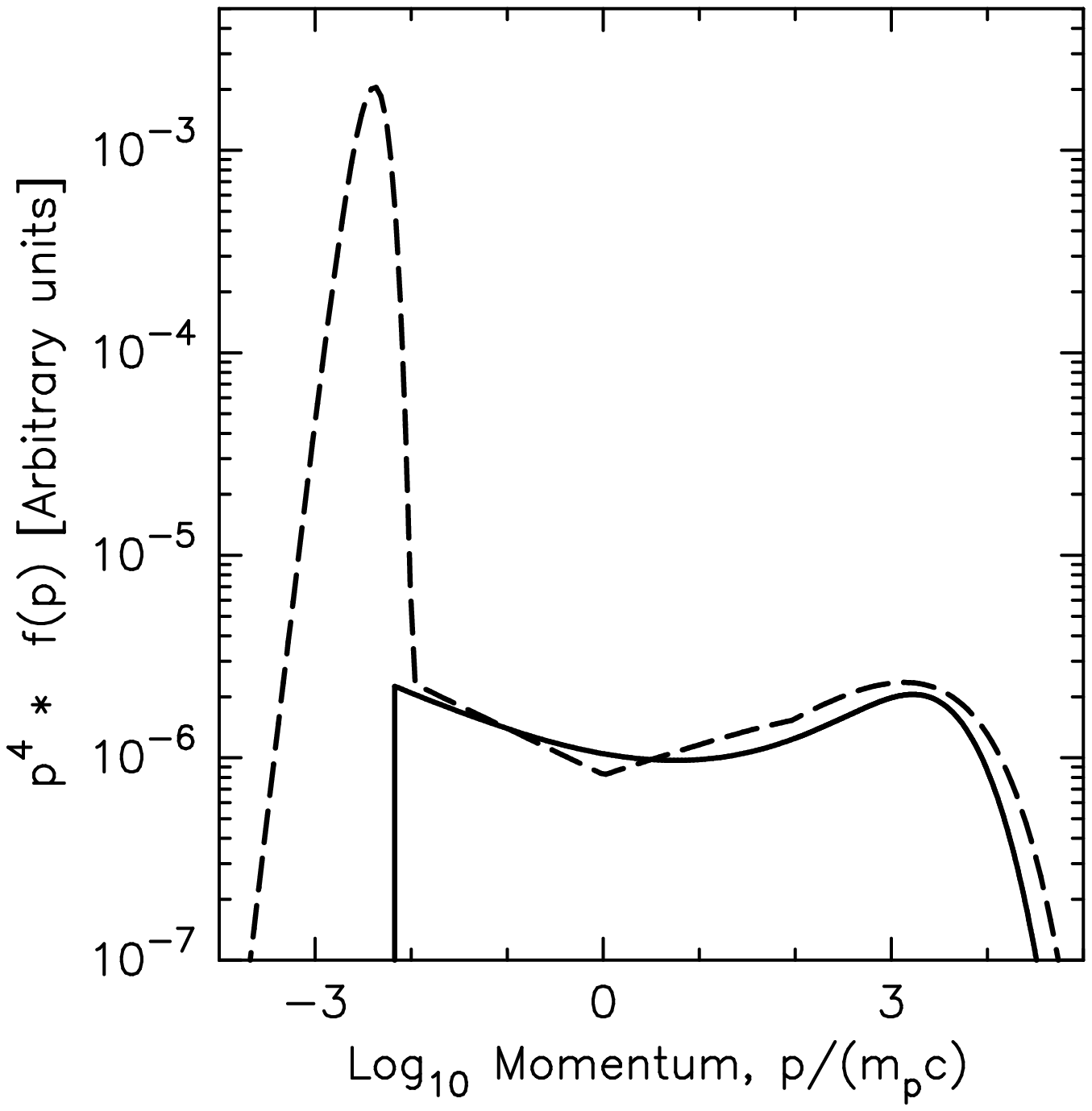}{
Proton phase space distributions from the simple model calculated here
(dashed line) and the kinetic model of Berezhko, Ksenofontov,
\& Petukhov (1999a) (solid line) for SN1006. The Berezhko
\etal\ result is the solid curve in their Figure~1d, and no adjustment
in normalization is made to improve the match.
\label{fig:bereSN1006}  }

\figureout{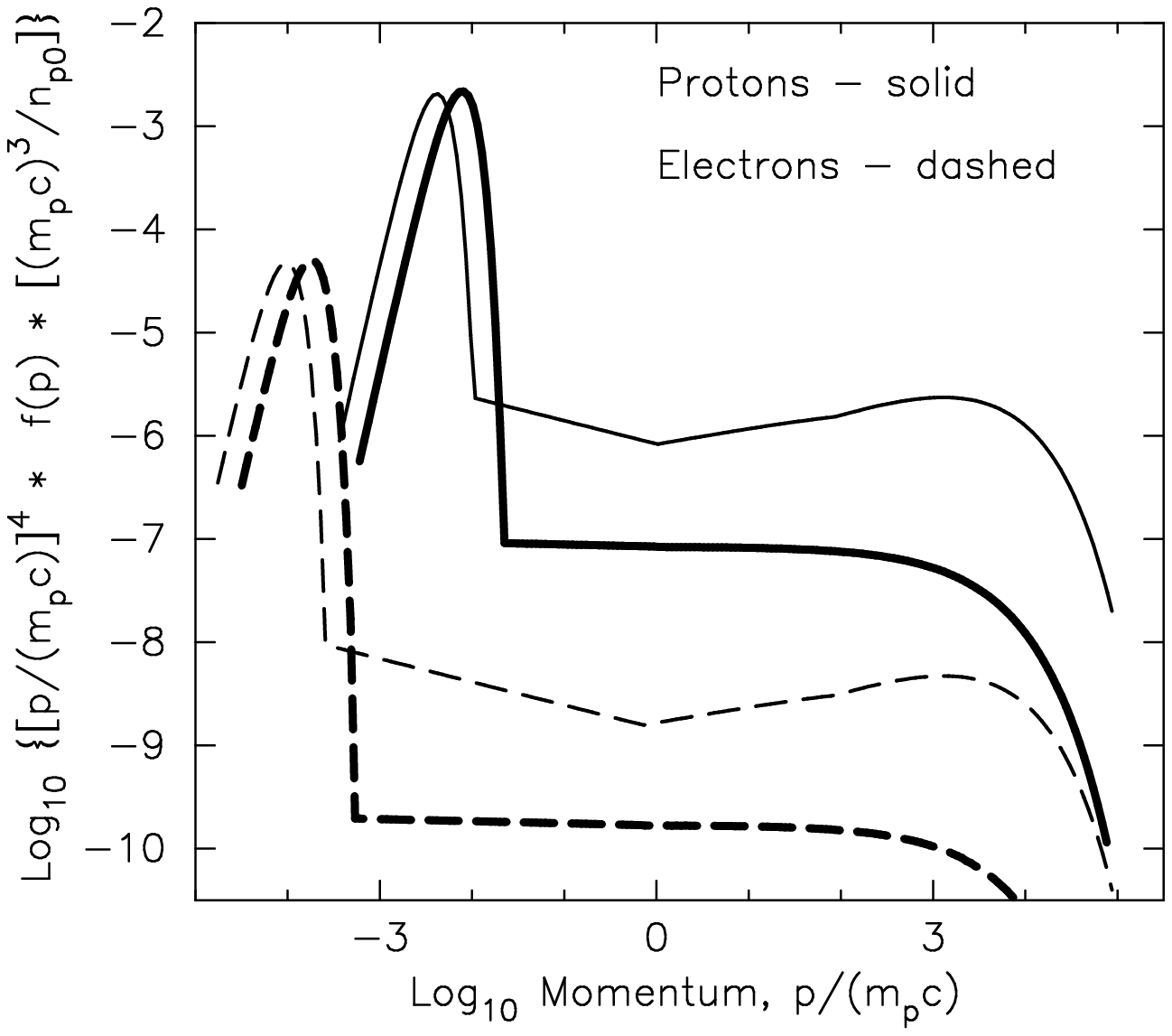}{
Electron and proton phase space distributions for the nonlinear and
test-particle models shown in Fig.~\ref{fig:SN1006}.  The light-weight
curves show the nonlinear $f(p)$, while the heavy-weight curves show
the test-particle $f(p)$. All input parameters, other than the proton
injection efficiency, are the same in the two cases. Compared to the
nonlinear result, the test-particle example has a higher shocked
temperature (as indicated by the position of the thermal peak) and a
lower shocked density. The difference in density is not apparent from
the figure because of the normalization.
\label{fig:testpart}  }


\begin{thebibliography}{99}
%
\bibitem[1994]{ABR94}
Achterberg, A., Blandford, R.D.,
Reynolds, S.P. 
\aaDCE{94}{281}{220}

\bibitem[1994]{Ammosov94}
Ammosov, A.E., Ksenofontov, L.T., Nikolaev, V.S., \& Petukhov, S.I.
\astroletsDCE{94}{22}{151}

\bibitem[1977]{ALS77}
Axford, W.I., Leer, E., \& Skadron, G.
\icrcplovdivDCE{11}{132}

\bibitem[2000]{Baring00}
Baring, M. G. 2000, in Proceedings of the Snowbird TeV Gamma-Ray Workshop,
eds. B. L. Dingus et al. (AIP, New York)

\bibitem[1997]{BaringEtal97}
Baring, M.G., Ogilvie, K.W., Ellison, D.C., \& Forsyth, R.J.
\apjDCE{97}{476}{889}

\bibitem[1999]{BaringEtal99}
Baring, M.G., Ellison, D.C., Reynolds, S.P., Grenier, I.A.,
\& Goret, P.
\apjDCE{99}{513}{311}

\bibitem[2000]{BJE2000}
Baring, M. G., Jones, F. C., \& Ellison, D. C. 2000, Ap. J. in press.

\bibitem[1978]{Bell78}
Bell, A.R.
\mnrasDCE{78}{182}{147}

\bibitem[1996]{Berez96}
Berezhko, E.G.
\appDCE{96}{5}{367}

\bibitem[1999]{BEapj99}
Berezhko, E.G., Ellison, D.C.
\apjDCE{99}{526}{385}

\bibitem[1996]{BereKrym88}
Berezhko, E.G., Krymsky, G.F.
\spuDCE{88}{31}{27}

\bibitem[1999]{BereKsen99}
Berezhko, E.~G., \& Ksenofontov, L.T.
\icrcsaltlakeDCE{4}{381}

\bibitem[1999a]{BereKP99a}
Berezhko, E.~G., Ksenofontov, L.T., \& Petukhov, S.~I.
\icrcsaltlakeADCE{4}{431}

\bibitem[1999b]{BereKP99b}
Berezhko, E.~G., Ksenofontov, L., \& Petukhov, S.~I. 1999b,
in press.
%

\bibitem[1996]{BerezEtal96}
Berezhko, E.G., Yelshin, V.K., \& Ksenofontov, L.T.
\JETPDCE{96}{82{\rm (1)}}{1}

\bibitem[1997]{BerezVolk97}
Berezhko, E.G., \& V\"olk, H.J. \appDCE{97}{7}{183}

\bibitem[1987]{BE87}
Blandford, R.D., \&  Eichler, D.
\phyreptsDCE{87}{154}{1}

\bibitem[1978]{BO78}
Blandford, R.D., \& Ostriker, J.P.
 \apjletDCE{78}{221}{L29}

\bibitem[2000]{Bykov2000}
Bykov, A.M., Chevalier, R.A., Ellison, D.C., \& Uvarov, Yu.A.
2000, \apjpressDCE

\bibitem[1982]{Chev82}
Chevalier, R.A. \apjDCE{82}{258}{790}

\bibitem[1996]{deJagerEtal96}
de Jager, O. C., Harding, A. K.,
 Michelson, P. F., Nel, H. I., Nolan, P. L.,
 Sreekumar, P., \& Thompson, D. J.
\apjDCE{96}{457}{253}

\bibitem[1997]{deJager97}
de Jager, O.~C., \& Mastichiadis, A. 
\apjDCE{97}{482}{874}

\bibitem[1991]{dorfi91}
Dorfi, E.~A. \aaDCE{91}{251}{597}.

\bibitem[1994]{Dorfi94}
Dorfi, E. A.
\apjsDCE{94}{90}{841}

\bibitem[1993]{DorfiB93}
Dorfi, E.A., \& B\"ohringer, H.
\aaDCE{93}{273}{251}

\bibitem[1996]{DorfiV96}
Dorfi, E.A., \& V\"olk, H.J.
\aaDCE{96}{307}{715}

\bibitem[1983]{Drury83}
Drury, L.O'C.
\rppDCE{83}{46}{973}

\bibitem[1994]{DAV94}
Drury, L.O'C., Aharonian, F.A., \& V\"olk, H.J.
\aaDCE{94}{287}{959} 

\bibitem[1981]{Eich81}
Eichler, D. \apjDCE{81}{247}{1089}

\bibitem[1984]{Eich84}
Eichler, D. \apjDCE{84}{277}{429}

\bibitem[2000]{ellison2000}
Ellison, D.C.
2000, Proc. of ACE workshop,
in press

\bibitem[1996]{EBJ96}
Ellison, D.C., Baring, M.G., \& Jones, F.C.
\apjDCE{96}{473}{1029}

\bibitem[1999a]{EBicrc99a}
Ellison, D.C., \& Berezhko, E.G.
\icrcsaltlakeADCE{4}{390}

\bibitem[1999b]{EBicrc99b}
Ellison, D.C., \& Berezhko, E.G.
\icrcsaltlakeBDCE{4}{446}

\bibitem[1984]{EE84}
Ellison, D.C., \& Eichler, D.
\apjDCE{84}{286}{691}

\bibitem[1990]{EMP90}
Ellison, D.C., M\"obius, E., \& Paschmann, G.
\apjDCE{90}{352}{376}

\bibitem[1998]{GPS97}
Gaisser, T.~K., Protheroe, R.~J., \& Stanev, T. \apjDCE{98}{492}{219}

\bibitem[1999]{Gehrels99}
Gehrels, N. \& Michelson, P. \appDCE{99}{11}{277}.

\bibitem[1993]{GBSE93}
Giacalone, J., Burgess, D., Schwartz, S.J., \&
Ellison, D.C.
\apjDCE{93}{402}{550}

\bibitem[1997]{GBSE97}
Giacalone, J., Burgess, D., Schwartz, S.J., Ellison, D.C., \&
Bennett, L. 
\jgrDCE{97}{102}{19,789} 

\bibitem[1981]{GoslingEtal81}
Gosling, J.T., Asbridge, J.R., Bame, S.J., Feldman, W.C., Zwickl,
R.D., Paschmann, G., Sckopke, N., \& Hynds, R.J.
\jgrDCE{81}{86}{547}

\bibitem[1998]{Green98}
Green, D.~A., A Catalogue of Galactic Supernova Remnants (1998 September
version), [{\tt http://www.mrao.cam.ac.uk/surveys/snrs/} ]

\bibitem[1983]{HSC83}
Hamilton, A.J.S., Sarazin, C.L., \& Chevalier, R.A.
\apjsDCE{83}{51}{115}

\bibitem[1991]{JE91}
Jones, F.~C., \& Ellison, D.~C. 
\ssrDCE{91}{58}{259}

\bibitem[1991]{KJ91}
Kang, H. \& Jones, T.~W. \mnrasDCE{91}{249}{439}

\bibitem[1995]{KJ95}
Kang, H. \& Jones, T.~W. \apjDCE{95}{447}{944}

\bibitem[1984]{KennelEtal84}
Kennel, C.F., Edmiston, J.P., Scarf, F.L., Coroniti, F.V.,
Russell, C.T., Smith, E.J., Tsurutani, B.T., Scudder, J.D., Feldman,
W.C., Anderson, R.R., Moser, F.S., \& Temerin, M.
\jgrDCE{84}{89}{5436}

\bibitem[1999]{kohnle99}
Kohnle, A. \etal\ \icrcsaltlakeDCE{5}{239}

\bibitem[1995]{Koyama95}
Koyama, K. \etal\ \natureDCE{99}{378}{255}

\bibitem[1977]{Krym77}
Krymskii, G.F. 1977,
{\itt Sov. Phys. Dokl.,} {\bff 22(6)}, 327

\bibitem[1996]{LamingEtal96}
Laming, J., M., Raymond, J.C., McLaughlin, B.M., \& Blair, W.P.
\apjDCE{96}{472}{267}

\bibitem[1982]{Lee82}
Lee, M.A.
\jgrDCE{82}{87}{5063}
                                             
\bibitem[1983]{Lee83}
Lee, M.A. 
\jgrDCE{83}{88}{6109}

\bibitem[1992]{Levinson92}
Levinson, A. 
\apjDCE{92}{401}{73}

\bibitem[1994]{Levinson94}
Levinson, A. 
\apjDCE{94}{426}{327}

\bibitem[1988]{Long88}
Long,, K.S., Blair, W.P., \& van den Bergh, S. \apjDCE{88}{333}{749} 

\bibitem[1997]{Lorenz97}
Lorenz, E. 1997, in Towards a Major Atmospheric \v{C}erenkov Detector,
   ed. O. C. de Jager (Wesprint, Pochefstroom) p.~415.

\bibitem[1997]{Malkov97}
Malkov, M.A. \apjDCE{97}{485}{638}

\bibitem[1999]{Malkov99}
Malkov, M.A. \apjletDCE{99}{511}{L53}

\bibitem[1996]{MdeJ96}
Mastichiadis, A. \& de Jager, O.~C. 
\aaDCE{96}{311}{L5}

\bibitem[1987]{Tang87}
M\"uller, D., \&  Tang, K.-K.
\apjDCE{87}{312}{183}

\bibitem[1995]{Mueller95}
M\"uller, D., \etal\
\icrcromeDCE{3}{13}

\bibitem[1996]{Reynolds96}
Reynolds, S.P.
\apjletDCE{96}{459}{L13}

\bibitem[1992]{RE92}
Reynolds, S.P., \& Ellison, D.C.  \apjletDCE{92}{399}{L75}

\bibitem[1979]{RL79}
Rybicki, G.B., \& Lightman, A.P. 1979,
``Radiative Processes in Astrophysics,'' 
John Wiley \& Sons, New York.
 
\bibitem[2000]{SKG2000}
Scholer, M., Kucharek, H., \& Giacalone, J.
\jgrpress2000DCE

\bibitem[1992]{STK92}
Scholer, M., Trattner, K.J., \& Kucharek, H.
\apjDCE{92}{395}{675}

\bibitem[1997]{SturnerEtal97}
Sturner, S.~J., Skibo, J.~G., Dermer, C.~D., \& Mattox, J.~R. 
\apjDCE{97}{490}{619}

\bibitem[1998]{Tanimori98}
Tanimori, T., \etal\ \apjletDCE{98}{497}{L25}

\bibitem[1999]{Terasawa99}
Terasawa, T., \etal\
\icrcsaltlakeDCE{6}{528}

\bibitem[1999]{Truelove99}
Truelove, J.K., \& McKee, C.F.
\apjsDCE{99}{120}{299}

\bibitem[2000]{Volk00} 
V\"olk, H.~J., \etal\ 2000, in Proceedings of
the Snowbird TeV Gamma-Ray Workshop, eds. B. L. Dingus et al.
(AIP, New York)

\bibitem[1999]{weekes99}
Weekes, T.~C., \etal\ 1999, VERITAS proposal
 [{\tt http://egret.sao.arizona.edu/vhegra/ vhegra.html}]

\bibitem[1996]{Zimm96}
Zimmermann, H.~U., Tr\"umper, J.~E., \& Yorke, H. 1996,
   R\"ontgenstrahlung from the Universe, MPE Report 263
   (Max-Planck-Institut f\"ur Extraterrestrische Physik, Garching)

\end{thebibliography}
\end{document}